\documentclass[lettersize,journal]{IEEEtran}

\usepackage[normalem]{ulem}
\usepackage{amsmath,amssymb,amsfonts}
\usepackage{caption}
\usepackage{graphicx}
\usepackage{textcomp}
\usepackage{xcolor}
\usepackage{mwe}
\usepackage{subfig}
\usepackage[ruled,vlined]{algorithm2e}
\usepackage{algpseudocode}

\usepackage{longtable}
\usepackage{array}
\usepackage[symbol]{footmisc}
\usepackage[normalem]{ulem}

\begin{document}
\let\WriteBookmarks\relax
\def\floatpagepagefraction{1}
\def\textpagefraction{.001}



\title{An Unsupervised Adversarial Autoencoder for Cyber Attack Detection in Power Distribution Grids}  


\author{{Mehdi~Jabbari Zideh, Mohammad Reza Khalghani, and Sarika Khushalani Solanki}


\thanks{M. Jabbari Zideh and S. K. Solanki are with the Lane Department of Computer Science and Electrical Engineering, West Virginia University, Morgantown, WV 26506, USA (e-mail: mehdijabbari@ieee.org, skhushalanisolanki@mail.wvu.edu)}

\thanks{M. R. Khalghani is with the Department of Electrical and Computer Engineering, Florida Polytechnic University, Lakeland, FL 33805, USA.
(email: mkhalghani@floridapoly.edu)}

\thanks{\emph{Corresponding author: Mehdi Jabbari Zideh.}}}
\maketitle
\begin{abstract}
Detection of cyber attacks in smart power distribution grids with unbalanced configurations poses challenges due to the inherent nonlinear nature of these uncertain and stochastic systems. It originates from the intermittent characteristics of the distributed energy resources (DERs) generation and load variations. Moreover, the unknown behavior of cyber attacks, especially false data injection attacks (FDIAs) in the distribution grids with complex temporal correlations and the limited amount of labeled data increases the vulnerability of the grids and imposes a high risk in the secure and reliable operation of the grids. To address these challenges, this paper proposes an unsupervised adversarial autoencoder (AAE) model to detect FDIAs in unbalanced power distribution grids integrated with DERs, i.e., PV systems and wind generation. The proposed method utilizes long short-term memory (LSTM) in the structure of the autoencoder to capture the temporal dependencies in the time-series measurements and leverages the power of generative adversarial networks (GANs) for better reconstruction of the input data. The advantage of the proposed data-driven model is that it can detect anomalous points for the system operation without reliance on abstract models or mathematical representations. To evaluate the efficacy of the approach, it is tested on IEEE 13-bus and 123-bus systems with historical meteorological data (wind speed, ambient temperature, and solar irradiance) as well as historical real-world load data under three types of data falsification functions. The comparison of the detection results of the proposed model with other unsupervised learning methods verifies its superior performance in detecting cyber attacks in unbalanced power distribution grids.
\end{abstract}



\begin{IEEEkeywords}
false data injection attacks, adversarial autoencoder, power distribution grids, unsupervised data-driven method, generative adversarial networks, cyber attack detection
\end{IEEEkeywords}

\maketitle

\section{Introduction}\label{Introduction}
\IEEEPARstart{I}{nstalling} advanced measurement devices such as smart meters, phasor measurement units (PMUs), and remote terminal units (RTUs) has facilitated the power system visibility, monitoring, and control \cite{protocols, cyber-physical, Dileep, mirzapour_GET}. The system operators can leverage this benefit to make crucial decisions in critical situations, especially in cases of cyber-power events. On the other hand, high-resolution measurement data transferred from the measurement units to the control centers offers this opportunity for attackers to intentionally threaten the security of the smart grids by creating malicious or unobservable attacks. This causes significant damage to the system's infrastructure or even cascading failures. Among the cyber attacks that are threatening cyber-power systems, false data injection attacks (FDIAs) have caught the attention of researchers due to their higher concealment to be detected by defense mechanisms \cite{sparsity_con, observer_bonab, vincent_cyberphysical}. In FDIAs, the attackers launch a malicious attack on the communication systems and manipulate the measurement data to make catastrophic consequences on major infrastructures of the power grids \cite{reachability_FDIA, masking}. However, the effects of such attacks will be mitigated by designing effective detection mechanisms.

Detection methods of cyber-attacks are broadly divided into three categories; 1) model-based algorithms \cite{survey_physics, model_based_AGC}, 2)  data-driven (machine learning) models\cite{survey_fdia, yashar_datadriven}, and 3) physic-informed machine learning methods \cite{PIML, PIConvAE}. Model-based approaches solely rely on the predefined mathematical representations of the system and need exact physics-based modeling to achieve accurate and reliable results. On the other side, data-driven models do not need an abstract model of the system and leverage machine learning (ML) strategies to capture hidden patterns and nonlinear characteristics of the system's measurements to effectively handle normal and abnormal data points. Integration of the mathematical relationships into the ML techniques creates physics-informed machine learning models. In the cyber-physical domains with dynamical representations, it is difficult or even infeasible to make mathematical modeling of the systems and one has to make specific assumptions to derive the physics-based information which may lead to low accurate results. Due to this reason and leveraging the extensive data collection from advanced metering devices, ML approaches have recently gained prominence in FDIA detection. The main focus of these research studies has been on FDIA detection in power transmission grids with balanced structure and load modelings. For instance, Authors in \cite{graph_FDIA} proposed a gated graph neural network based on an attention mechanism to assign weights to systems nodes for detecting FDIAs in power transmission grids. A defense mechanism based on interval state predictor is presented in \cite{DL_state_prediction} to formulate the variation bounds of system variables as an optimization method. A deep belief network was also applied for extracting the nonlinear features and improving the detection accuracy of the defense mechanism. Goyel $et$ $al.$ \cite{Data_Integrity} proposed an attack generation framework in which the attacker only has limited information about the power grid. They also proposed an ensemble learning algorithm for data integration attack detection. Ref \cite{multivariat_eensemble} applied multivariate ensemble classification by combining the extreme learning machine, light gradient enhancement machine, and extreme gradient enhancement to identify traces of FDIAs in cyber-physical energy systems. Authors in \cite{realtime} proposed a Conditional Deep Belief Network (CDBN) to extract and analyze temporal patterns of FDIAs presented in the sensor measurement data in DC power distribution grids. 
These are examples of ML applications for FDIA detection in transmission grids. While the detection of FDIAs is an important task for securing smart grids against cyber threats, the equally crucial task is the development of robust techniques to recover true values of measurements and system states after cyber-attack occurrence \cite{inertia_recovery, generalized_recovery, gan_recovery}. The work in \cite{inertia_recovery} and \cite{generalized_recovery} utilized the inertia effect of measurement data and imposed secure and normal operation bounds as constraints to revert state variables to the pre-attack values. In \cite{gan_recovery}, a generative adversarial network (GAN) model is integrated with a physical model to generate ideal non-tempered data and replace manipulated measurements for recovering the manipulated state estimation data.

Data-driven approaches are preferred over model-based models, such as state estimation methods, for FDIA detection in power distribution systems. The research study in \cite{DL_cyber_review} presents a thorough overview of the applications of deep learning (DL) models in smart grids. It discusses different DL structures including convolutional neural network (CNN), restricted Boltzmann machine (RBM), recurrent and residual neural networks, graph neural networks, and GANs as well as their applications for electricity theft attacks, FDIAs, false command attacks, communication traffic and adversarial learning attacks. The main features of distribution systems rest on the dynamic nature of distributed generation (DG) and load variations, and more importantly, the unbalanced configuration of the grids that make them more complicated than the transmission systems \cite{DG_impact, pf_tiwari}. Therefore, developing an accurate model of the system is not always feasible. Another challenge that researchers deal with for anomaly detection in power distribution grids is the scarcity of labels, primarily due to security concerns. Therefore, they need to manually label the data samples which is time-consuming, prohibitively expensive, and demands a high level of expertise in the cyber-power domain \cite{transfer_hai, AD_building}. The limited number of labeled data restricts the generalization of the methods that are highly dependent on the labeled data samples (supervised and semi-supervised learning methods). As a result, unsupervised learning models have found more practical applications in scenarios of scarce labeled data. 

The research studies mostly used unsupervised methods for detecting physical events with specific characteristics in power distribution systems, such as the works in \cite{aligholian2021, fast_ramped, aligholian2020, solar_farm, david2020}. However, only a small number of researchers have utilized these methods for the detection of cyber-attacks with unknown behavior and patterns in power distribution grids with unbalanced structures. 
The applied ML methods are mostly supervised or semi-supervised. For instance, authors in \cite{semi-Supervised} utilized an autoencoder and integrated it into generative adversarial networks (GANs) to detect unobservable attacks in three-phase unbalanced distribution systems. 
The proposed semi-supervised method relies on the dimensionality reduction of measurements and extracts the main features of the partially-labeled data to identify stealthy outliers or attacks. 
In \cite{hierarchical}, a multi-layer long short-term memory (MLSTM) is applied to an adaptive hierarchical framework to detect and localize real-time cyber attacks in DER-integrated distribution systems based on the online waveform measurement data.
Authors in \cite{volt_reg} presented a two-stage ML approach for data falsification attack detection and localization. They employed random forest (RF) to forecast voltage magnitudes based on historical voltage and meteorological data. 
The forecasted voltage measurements are compared with the actual measured data using a logistic regression model in the second stage to identify the falsified voltages.
Authors in \cite{coordinated} presented CNN, multi-layer perception (MLP), and residual neural network (ResNet) to detect falsified measurements of wind turbines and PVs. 
They introduced additive and deductive attack mechanisms and compared the detection performance of the neural networks with classical ML algorithms.
Naderi $et$ $al.$ \cite{naderi2022} proposed an MLP network to detect FDIAs in distribution grids. The proposed method utilizes the historical load data and weather data to predict the measurements for the next time interval, and by identifying the relationships between the measurements, it can warn the data manipulation. 
A supervised NN-based model is presented in \cite{raghuvamsi} by combining temporal CNN with an attention mechanism to create a convolutional denoising autoencoder for FDIA location and missing data detection. The proposed approach utilizes the advantages of dimensionality reduction, the extraction of spatiotemporal correlation, and the correlation among the measurement samples to reconstruct the input data for the recovery of attacked data.
A binary classification method using real measurements is implemented in \cite{radhoush2023} to detect FDIA. In the proposed methodology, FDIA detection and distribution system state estimation are separately and simultaneously applied to {\textmu}PMU measurements to identify the falsified measurements.

The previously mentioned research studies employed supervised or semi-supervised learning methods, requiring the label of the training data.  
Researchers have recently developed new unsupervised learning models for anomaly detection. Authors in \cite{super_resolution_perception} proposed a spatiotemporal graph DL-based (STGDL) framework to detect cyber attacks and employed a super-resolution perception (SPR) network to reconstruct the high-frequent state estimation data to improve the learning performance of the STGDL model. A convolutional autoencoder was developed in \cite{Conv_AE_ehsani} to detect anomalies in distribution PMUs. It employs feature extraction and obtains essential information from the measured data for anomaly detection.
To address challenges encountered by supervised or semi-supervised learning models, this paper proposes an adversarial autoencoder (AAE) leveraging the benefits of autoencoders for data reconstruction and generative adversarial networks (GANs) for capturing the hidden distribution of data measurements to detect FDIAs in unbalanced power distribution systems. We apply long short-term memory (LSTM) in the structure of the encoder and decoder to help autoencoders capture the spatiotemporal correlations of the measurement data. The generative model utilizes two discriminators, one for latent space and the other for reconstructed data, where CNN is employed for each of them to identify the hidden patterns of the generated data.

The contributions of this work can be summarized as follows:
\begin{enumerate}
\item{This paper proposes an unsupervised AAE learning-based method for unobservable FDIA detection in unbalanced power distribution grids. The developed method can successfully capture the nonlinear characteristics of the measurements to detect different types of FDIAs (deductive, additive, and combined attacks) in three-phase unbalanced power distribution grids without the need for knowledge of the system and labeled data. This feature makes this method very powerful to be implemented in real-world cases with unbalanced configurations and nonlinearity of DER generations.}
\item{We test the performance of the proposed unsupervised method under three attack scenarios using IEEE 13-bus and 123-bus systems. To consider the uncertainties in the distribution systems, we use realistic meteorological data for the generation of PVs and wind turbines, along with real historical load data. Finally, the detection results of this method are compared to other data-driven models to show its superiority in detecting anomalous scenarios.}
\end{enumerate}

The remainder of this paper is organized as follows. Section \ref{FDIA_Development} introduces the data falsification attack mechanisms and three FDIA functions. Section \ref{AAE_model} presents data preprocessing and modeling before training and testing of the proposed model as well as the architecture of the detection framework and the proposed AAE model for FDIA detection. The performance of the proposed AAE model for detecting FDIAs in unbalanced power distribution grids is evaluated in Section \ref{Results}. Finally, Section \ref{conclusion} concludes this paper.

\section{Development of FDIAs in Power Distribution Systems}\label{FDIA_Development}
The measurements collected from supervisory control and data acquisition
(SCADA) systems and micro-PMUs, which include current, voltage, and power flow measurements with high sampling rates, are utilized for wide-area monitoring of the real-time operation of smart power distribution systems. These measurements transmitted through the communication networks are vulnerable to malicious attacks. The attackers aim to manipulate the measurements without being detected with the limited information they acquire through historical data. In this paper, three types of FDIAs are considered \cite{volt_reg, camouflage}: a) deductive attacks, b) additive attacks, and c) camouflage (combined attacks of deductive and additive).

Let $z$ = \{$z_1, z_2, ...., z_m$\} be a series of $m$ measurement data reported by an installed sensor in the distribution network. For normal or unbiased measurements, the reported data is the same as the actual measurement, while the compromised measurements $z^{att}$ can be modeled through one of the following models.

\subsection{Deductive Attacks}\label{deductive}
In this type of attack, the attacker tries to report the measured values below the actual measurements by deducting some amount $\Delta z$; therefore, the attacked measurement $i$ is modeled as
\begin{equation}\label{deductive_Eq}
z^{att}_i = z_{i}-\Delta z_i 
\end{equation}
where $\Delta z_i$ denotes a percentage of the real measurement $z_i$. This attack type typically occurs when the real measurements are above the allowed threshold. It is to deceive the controllers that the measurements are within the limit and prevent them from taking appropriate corrective actions.

\subsection{Additive Attacks}\label{additive}
In this type of data falsification attack, the adversary manipulates the measurements to be reported as normal data while the actual measured values are below a certain threshold. The manipulation is implemented through constant and time-dependent partial increments. The model for additive manipulation is defined as
\begin{equation}\label{additive_Eq}
z^{att}_i = z_{i}+\Delta z_i 
\end{equation}

\subsection{Combined (camouflage) Attacks}\label{camouflage}
In combined attacks, data falsification is implemented by launching deductive attacks on certain portions of the measurements while the other parts are compromised by additive manipulation. For the combined attacks, we assume that out of the total number of compromised data, half of them are manipulated by additive and the other half by deductive attacks. It is also assumed that the attacker has the minimum information about the system's architecture and parameters required for launching the attacks. 

For successful FDIA to bypass the bad data detector (BDD) mechanism, the state estimation (SE) criteria by the traditional BDD should be met. In the power system SE framework, the nonlinear function $h$ represents the relationships between the system measurement $z$ and system states $y$ by $z=h(y)$. In the traditional BDD mechanism, if the residual of SE; $\|r=z-h(\hat{y})\|_2$ is greater than a predefined threshold $\tau$, where $\hat{y}$ is the estimated state, the set of measurements are flagged as anomalies. From the attacker's perspective, the falsified measurement $z^{att}$ needs to satisfy $\|z^{att}-h(\hat{y})\|_2 \leq\tau$ to be undetectable. We have set the manipulated measurements to be within 5\% of actual measurements to make them challenging for the BDD mechanism to detect.

\section{Proposed Unsupervised FDIA detection Approach} \label{AAE_model}
This section presents data processing and modeling for preparing the measurements for training and testing and the deep learning-based algorithm for unobservable FDIA identification in unbalanced three-phase power distribution grids integrated with distributed energy resources (DERs).

\subsection{Data Preprocessing}
Data preprocessing and modeling prior to NN training are crucial for obtaining the desired outcomes. As raw data is typically prone to missing values, noise, or inconsistency, its quality needs to be improved to get more accurate results. There are several techniques for preparing data before feeding them to neural networks such as feature extraction \cite{solar_irad}, data normalization \cite{bhanja}, and data sub-sampling \cite{sliding_window}. In the following, data normalization and data sub-sampling are described and used for preparing the training and test data to train the proposed AAE model. 
\begin{figure}[t!]
    \centering
    \includegraphics[clip,trim=4cm 0.9cm 1.1cm 4.5cm, width=1.35\linewidth]{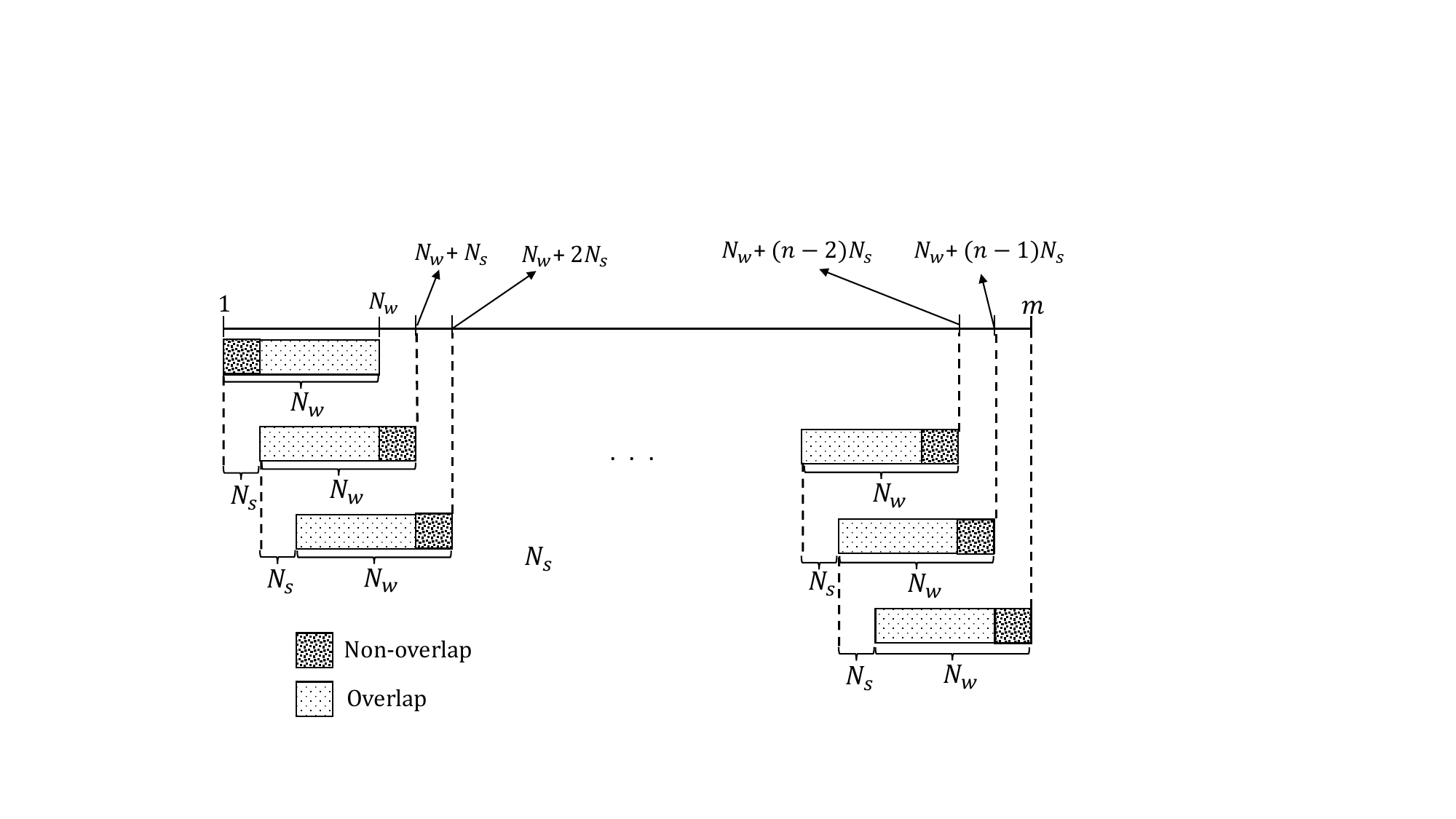}
\vspace*{-8mm}
 \caption{\small  Rolling-based sliding window strategy for input dataset. }
 \label{fig:sub-sampling}
\end{figure}
\subsubsection{Data Normalization}
Data normalization as a basic technique for data pre-processing is applied to produce high-quality data \cite{stock_idex, nemati_acoustic}. There are several benefits for data normalization before training DNN models. It brings the input datasets to a similar scale and helps the ML algorithms to be trained faster with a more stable and robust learning process. Moreover, it prevents the features with larger scales from overshadowing the other ones making the gradient update more effective. In addition, the regularization effect on the input data helps the NN models to better generalize to data outside the training window. There are several data normalization techniques such as min-max, decimal scaling, $z$-score, median, Sigmoid, and Tanh normalization techniques \cite{jayala}. In this paper, we use the $min-max$ normalization technique. In this technique, the data inputs are scaled into a predefined range of [-1,1] or [0,1]. The main advantage of this technique is that it preserves the important relationships and distances between measurements highlighting the effects of outliers or anomalies which is essential for anomaly detection. Assume that $x_{min}$ and $x_{max}$ are the minimum and maximum of dataset $X$, respectively. The data point $x$ is normalized to $x_{norm}$ in the range [low, high] as follows.
 \begin{equation}\label{normalization}
 \vspace{-6mm}
x_{norm}=low + \frac{(x-x_{min})}{(x_{max}-x_{min})}*(high - low)
\vspace{3mm}
\end{equation}

\subsubsection{Data Sub-sampling}
To obtain the training dataset, we use a rolling-based sliding window strategy to divide the whole dataset into sub-datasets with window size $N_w$ and step size $N_s$. This strategy divides the original dataset $X$ with size $m$ ($X={(x_1, x_2, ...., x_m)}$) into $n$ equally-sized samples ($n = \frac{m-N_w}{N_s}$) as shown in Fig. \ref{fig:sub-sampling}. For each of two consecutive windows, there is an overlap with size $N_w-N_s$ to capture variation trends and temporal dependencies in the dataset. The non-overlapping part is the new data with size $N_s$ which is called the step size.

\begin{figure*}
    \hspace{0mm}
    \centering
    \includegraphics[clip,trim=2.27cm 5cm 2.85cm 3cm, width=0.99\textwidth]{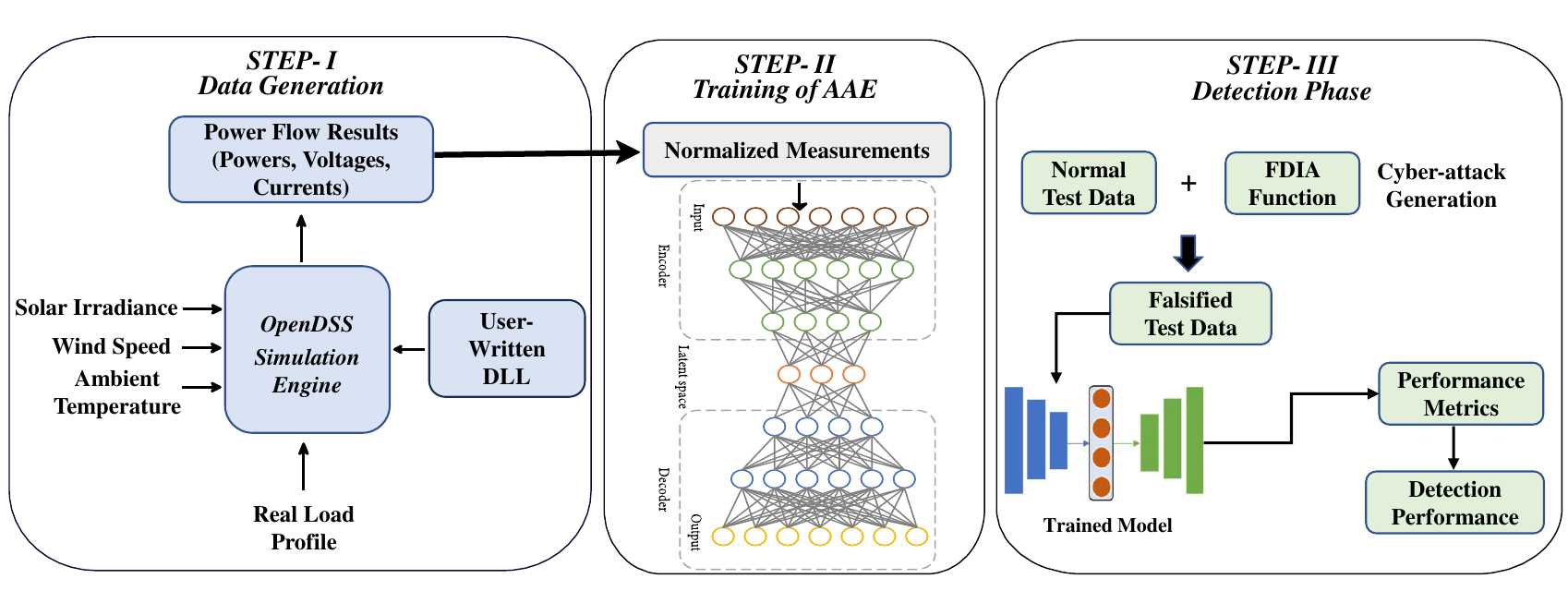}
    \caption{\small The proposed comprehensive framework for FDIA detection}
    \label{fig:methodology}
\end{figure*}

The proposed generative model, which is based on the concept of the model presented in \cite{TadGAN}, utilizes an autoencoder as the generator of GANs to reconstruct the normal measurements, including three-phase currents, voltages and power flow of the lines. The discriminator of the GANs model aims to distinguish between the reconstructed data and real measurements.  The comprehensive detection mechanism is shown in Fig. \ref{fig:methodology}. We present a detailed description of the proposed framework for FDIA detection including the architecture of GANs and autoencoder as follows.

\subsection{Autoencoder}\label{Autoencoder}
\begin{figure*}
    \centering
      \includegraphics[clip,trim=2.5cm 6cm 1.4cm 1.3cm, width=0.9\linewidth]{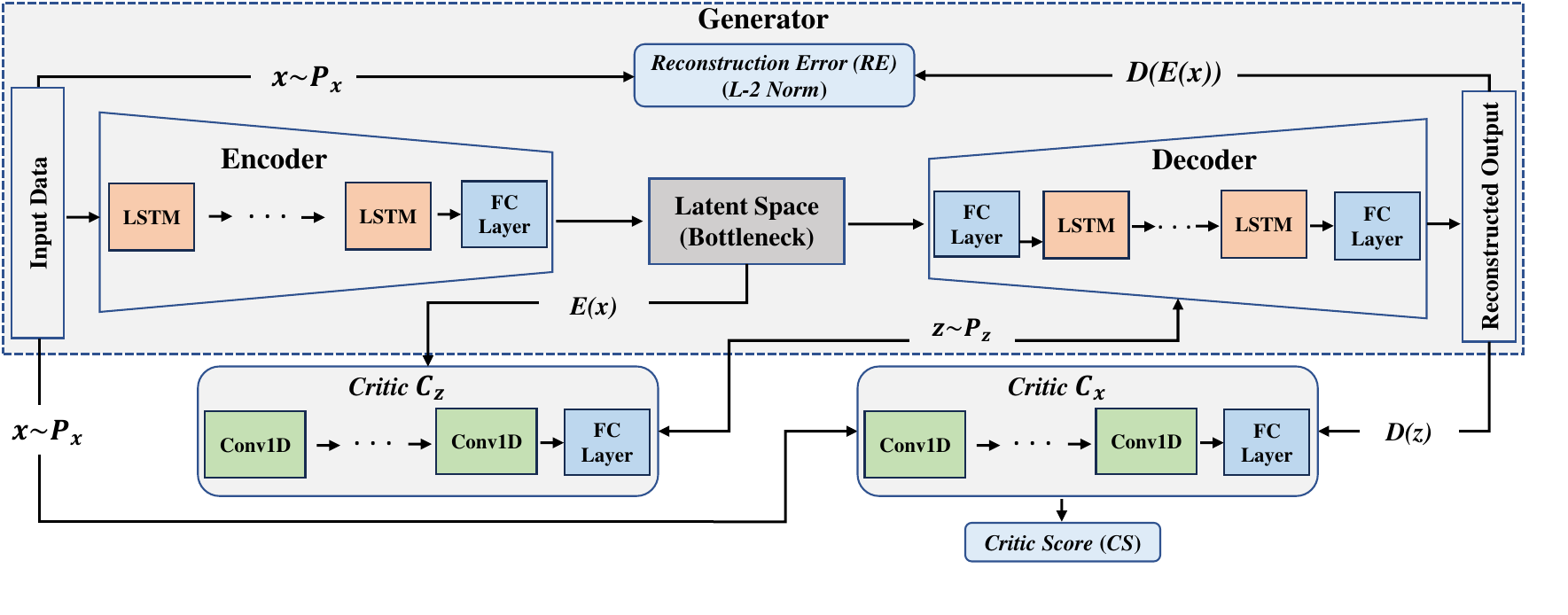}
 \vspace*{-4mm}
 \caption{\small Architecture of Adversarial Autoencoder.}
 \label{fig:Adversarial Autoencoder}
 \vspace*{-4mm}
\end{figure*}

The main idea for using reconstruction-based methods such as autoencoders for anomaly detection is that since anomalous data lose information after transferring to a lower dimensional space, their reconstructed outputs have a large reconstruction error, which is a sign of flagging anomalies \cite{autoencoder_AD}. Therefore, an effective model cannot reconstruct abnormal data points. Fig. \ref{fig:Adversarial Autoencoder} shows the proposed AAE architecture in which the autoencoder acts as the generator of the GAN model.
It learns two mappings; the encoder $\mathcal{E}$ maps the time-series data from domain $\mathcal{X}$ to a lower dimensional space $\mathcal{Z}$ while the decoder $\mathcal{D}$ does a reverse action to reconstruct the actual input data. The generative model applies the autoencoder to reconstruct the time-series data by minimizing the following loss function:
\begin{equation}\label{AE}
\mathcal{L_{AE}}\left( \theta_{\mathcal{E}}, \theta_{D} \right) =\frac{1}{m}\sum_{k=1}^{m}\left| x_{k}-x^{rec}_{k} \right|^{2} +\lambda \mathcal{R}\left( \theta \right)
\end{equation}
where $\theta_{\mathcal{E}}$ and $\theta_{\mathcal{D}}$ represent the parameters of encoder and decoder, respectively, $m$ is the data sample size, $x$ is the actual data sample, $x^{rec}$ is the reconstructed value, $\lambda$ is a hyper-parameter preventing the weights from overgrowing, and $\mathcal{R}$ is regularization which prevents the NN models from overfitting. It is generally added as a constraint to the loss function to help models learn more general and important patterns that better fit with general unseen data. To reconstruct more realistic data, L2 (Lasso) regularization has been used that applies the squares of the training weights to the loss function to make some of them exactly zero.

Autoencoders are known for their ability to capture the nonlinear behavior of high dimensional data \cite{AD_Yin}. These high-dimensional data are transferred to a lower-dimensional space using an encoder to extract high-level nonlinear spatial features. These features are subsequently employed for the data reconstruction. The strength of autoencoders to model nonlinearities in the data stems from the following aspects. First, by using multiple hidden layers in the structure of the encoder and decoder, they can compress the high-resolution data to latent-space representations and learn nonlinear and complex transformations. Secondly, the use of nonlinear activation functions in the hidden and output layers of the autoencoder helps the model learn complex mappings and approximate any continuous functions. Thirdly, similarly to other deep learning models, regularization techniques such as dropout can be applied to these models to improve their capabilities in handling the nonlinearities and randomness in complex uncertain data. These features make autoencoders to be applied in unbalanced power distribution systems with nonlinear characteristics of loads and DER generation. We use LSTM in the structure of the encoder and decoder to capture the temporal dependencies of time-series data. The use of LSTM in the structure of the autoencoder will make the model powerful at capturing the spatiotemporal correlations in the sliding windows of the measurements. The details of the structure and parameters of the autoencoder are presented in Section \ref{parameters}.
\subsection{GAN}
The GAN model consists of two neural networks; a generator and a discriminator, that compete with each other and play a min-max game to achieve a Nash equilibrium point \cite{iterative} for generating high-quality outputs that resemble the input data. In the unsupervised GAN model, the generator draws inputs from the distribution of input training data and aims to generate outputs that approximate this distribution as closely as possible\cite{GAN}. On the other hand, the goal of the discriminator is to distinguish the generated (synthetic) data generated by the generator and the original (authentic) training data. We utilize Wasserstein loss \cite{WGAN} to train the GAN model, where both discriminators try to minimize the maximum Wasserstein-1 distance between actual and generated data samples. We have two adversarial training procedures, one between the encoder $\mathcal{E}$ and critic $\mathcal{C_z}$, and the other between the decoder $\mathcal{D}$ and critic $\mathcal{C_x}$. For mapping $\mathcal{X} \rightarrow \mathcal{Z}$, the objective function is as follows.
\begin{equation}
\label{Critic_z}
    \begin{split}
        \mathbf{\underset{\theta_{\mathcal{E}}}{min}}\ \mathbf{\underset{\theta_{\mathcal{C}_z}}{max}}\ \mathcal{L}_z(\mathcal{C}_z , \mathcal{E}) & = \mathbf{E}_{z\sim \mathcal{P}_z}[\mathcal{C}_z(z)] \\
        & - \mathbf{E}_{x\sim \mathcal{P}_x}[\mathcal{C}_z(\mathcal{E}(x))]
    \end{split}
\end{equation}
where $\mathcal{P}_x$ and $\mathcal{P}_z$ denote the distributions of real training data and random normal distribution ($z\sim \mathcal{P}_z$ = $N(0,1)$), respectively.

Following the same methodology, the mapping $\mathcal{Z} \rightarrow  \mathcal{X}$ is implemented with the following objective function:
\begin{equation}
\label{Critic_x}
    \begin{split}
        \mathbf{\underset{\theta_{\mathcal{D}}}{min}}\ \mathbf{\underset{\theta_{\mathcal{C}_x}}{max}}\ \mathcal{L}_x(\mathcal{C}_x , \mathcal{D}) & = \mathbf{E}_{x\sim \mathcal{P}_x}[\mathcal{C}_x(x)] \\
        & - \mathbf{E}_{z\sim \mathcal{P}_z}[\mathcal{C}_x(\mathcal{D}(z))]
    \end{split}
\end{equation}
For both critics, the gradient penalty regularization is applied to control the weights and impose penalties on the gradients that are not equal to one.
\subsection{Adversarial Autoencoder}

The AAE model leverages the features of the autoencoder to encode and decode time-series measurements and applies the two critics to distinguish between fake and real measurements. Assume that the training dataset $X$ consists of $m$ unlabelled measurements ($x_1, x_2, ...., x_m$). The encoder $\mathcal{E}$ is trained to encode the input data to lower dimensional space with a prior distribution $\mathcal{P}_{latent}$ while the decoder generates the output time-series data ($x^{rec}_1, x^{rec}_2, ...., x^{rec}_m$) with the probability distribution $\mathcal{P}_{model}$ to be as similar as possible to the distribution of input data $\mathcal{P}_{x}$. To train the encoder and decoder as the components of the GAN's generator, two critics, $\mathcal{C}_z$ and $\mathcal{C}_x$ are employed, respectively. The goal of Critic $\mathcal{C}_z$ is to evaluate the quality of the time-series data in the latent space while the critic $\mathcal{C}_x$ is trained to distinguish the real input data from the reconstructed outputs of the decoder.

To reconstruct the measurement data using the proposed AAE method, all of the objective functions in Eqs. \ref{AE}, \ref{Critic_z}, and \ref{Critic_x} are combined to create a min-max problem as follows.
\begin{equation}\label{AAE}
    \begin{split}
        \mathbf{\underset{{\theta_{\mathcal{E}},\theta_{\mathcal{D}}}}{min}}\ \mathbf{\underset{{\theta_{\mathcal{C}_z},\theta_{\mathcal{C}_x}}}{max}}\ \mathcal{L_{AE}}\left( \theta_{\mathcal{E}}, \theta_{\mathcal{D}} \right) + \mathcal{L}_z(\mathcal{C}_z , \mathcal{E}) + \mathcal{L}_x(\mathcal{C}_x , \mathcal{D})
    \end{split}
\end{equation}
where the first loss term ensures the individual mapping from the input space to the reconstructed space, while the two other losses guarantee the feasibility of K-Lipschitz functions to avoid gradient explosion and stable training of the model.

Algorithm 1 shows the pseudo-code for the training and testing of the proposed AAE model. In this algorithm, $gp$ is a gradient penalty with a penalty coefficient $\lambda$ to encourage the gradients to go towards 1 \cite{gradient}.

\section{Experimental Results} \label{Results}
\begin{figure}
    \centering
    \vspace*{-44mm}
    \includegraphics[clip,trim=2.1cm 0.5cm -2.5cm 0cm, width=1.25\linewidth]{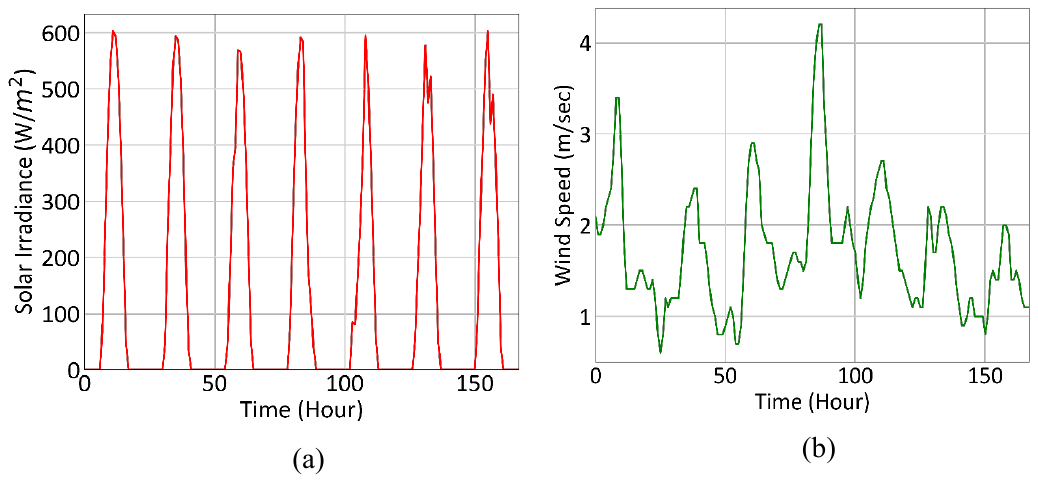}
\vspace*{-55mm}
 \caption{\small  Profiles for San Diego during the first week of January 2021: (a) solar irradiance, (b) wind speed.}
 \label{fig:profiles}
\end{figure}

\begin{figure}[t!]
    \subfloat[]{%
        \includegraphics[width=.52\linewidth]{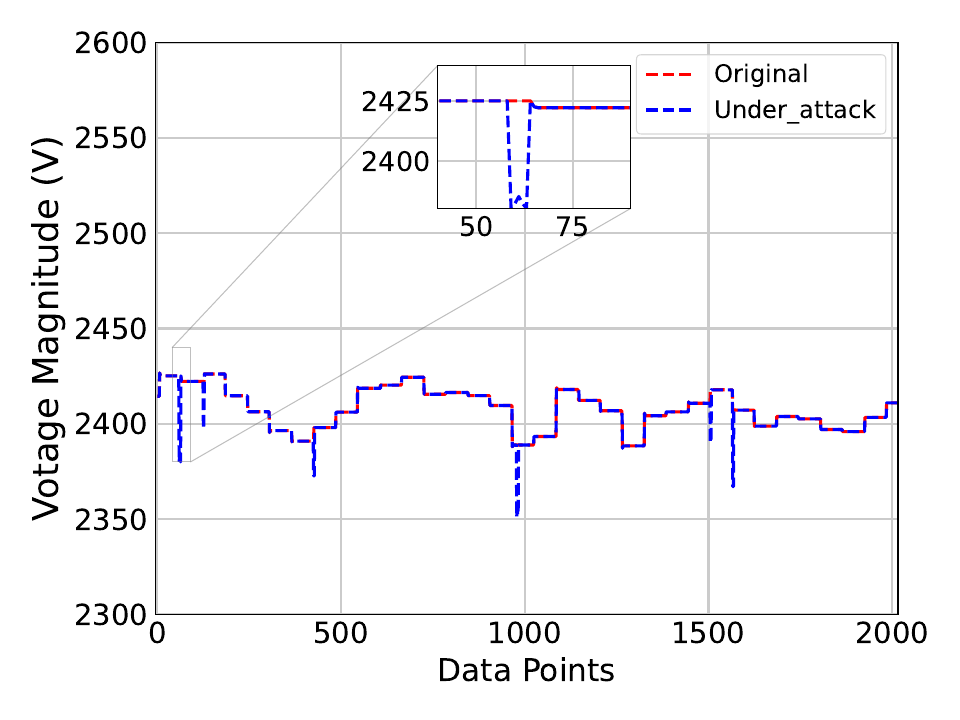}%
        \label{deductive}%
    }
    \subfloat[]{%
        \includegraphics[width=.52\linewidth]{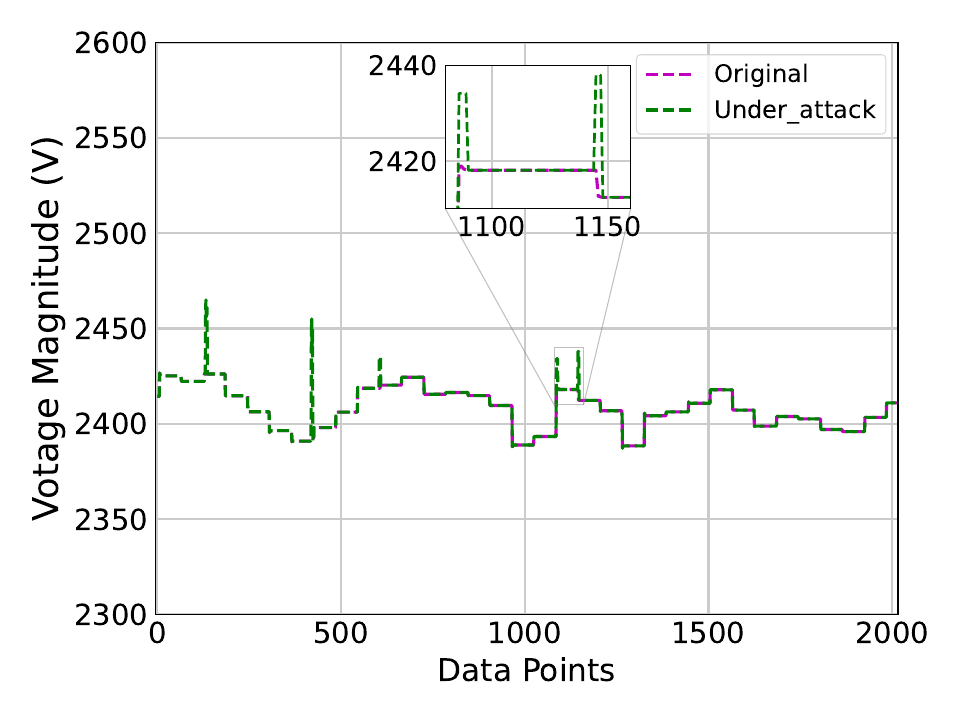}%
        \label{additive}%
    }
    \caption{Power flow result (voltage magnitude) at bus 35 for IEEE 123-bus: (a) under normal and deductive attacks, (b) under normal and additive attacks.}
    \label{deductive_additive}
\end{figure}

To investigate the applicability and performance of the proposed AAE method for FDIA detection in unbalanced power distribution systems, we test it in IEEE 13-bus  and 123-bus systems \cite{test_systems}. To incorporate the randomness and uncertainties of real-world cases into these grids, PV and wind generation units are integrated into them. We use the real-world historical data for solar irradiance, ambient temperature, and wind speed for the city of San Diego in the first week of 2021 \cite{NREL}. Fig. \ref{fig:profiles} shows the profiles of wind speed and solar irradiance during the aforementioned period. The system loads are modeled based on realistic load data patterns of San Diego Gas and Electric (SDGE) during the same period. Moreover, multiple measurement units are located in different nodes and lines to measure the voltages, currents, and power flows of the two systems. Table \ref{tbl:System_Details} summarizes the details of the DG and monitor placement for both systems.
The simulation for generating measurements, including active and reactive power, currents, and voltages, is implemented by the open-source distribution system simulator (OpenDSS) software. OpenDSS was developed by the Electric Power Research Institute (EPRI) to conduct grid integration of DERs and perform complex and customized simulations for the distribution grids. We have simulated the test case for a one-week duration with one-minute time intervals to collect 10080 measurement samples. The test samples are manipulated by the attack types mentioned in Section \ref{FDIA_Development} to be used for model evaluation. Fig. \ref{deductive_additive} shows the power flow result (voltage magnitude) at bus 35 in the IEEE 123-bus system under normal and attack scenarios.
\begin{table}[b]
    \centering
    \vspace*{-10mm}
    \caption{\small  DGs and Monitors Locations in Test Systems.}
      \includegraphics[clip,trim=3.2cm 17.3cm -1.8cm 2.53cm, width=1.25\linewidth]{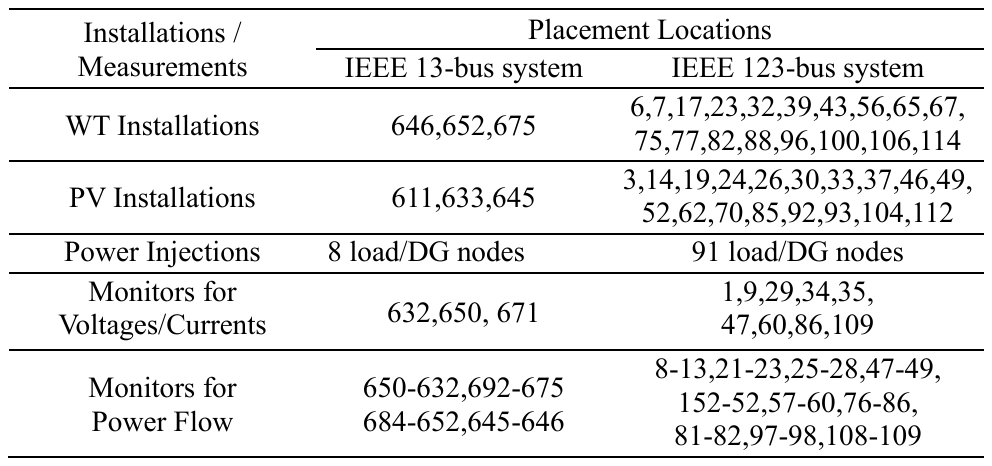}
 \label{tbl:System_Details}
\end{table}

\IncMargin{1em}
\begin{algorithm}[htbp]\label{alg1}
\SetKwData{Left}{left}
\SetKwData{This}{this}
\SetKwData{Up}{up}
\SetKwFunction{Union}{Union}
\SetKwFunction{FindCompress}{FindCompress}
\SetKwInOut{Input}{Input data}\SetKwInOut{Output}{Required}\SetKwInOut{Initial}{Initialization}
\SetKwComment{comment}{\#}{}
\Input{Training and Test Data ($X_{train}$, $X_{test}$)}
\Output{$Epoch$: number of epochs, $b$: batch size\\$N_{w}$: window size, $\alpha$: learning rate\\$N_c$, number of iteration of the critics per epoch}
\Initial{Weights $w_{\mathcal{C}_x}$, $w_{\mathcal{C}_z}$, $w_{\mathcal{E}}$, $w_{\mathcal{D}}$}
\BlankLine

\textbf{Training:}\\
\For{each $Epoch$}{
    Sample {$(x_i^{1...N_w})_{i=1}^b$} from training data\\
    Sample {$(z_i^{1...N_w})_{i=1}^b$} from random data\\
    \For{$l=1,..., N_c$}{
    \textbf{Critic} \textbf{$\mathcal{C}_x$:}\\
    Fix weights $w_{\mathcal{C}_z}$, $w_{\mathcal{E}}$, $w_{\mathcal{D}}$\\
   $ \mathcal{L}_{\mathcal{C}_x}=\frac{1}{b}\sum_{i=1}^{b} \mathcal{C}_x(x_i)-\frac{1}{b}\sum_{i=1}^{b} \mathcal{C}_x(\mathcal{D}(z_i))+\lambda*gp(x_i,\mathcal{D}(z_i))$\\
   $w_{\mathcal{C}_x}=w_{\mathcal{C}_x}-\alpha \sum_{x}\nabla_{w_{\mathcal{C}_x}} \mathcal{L}_{\mathcal{C}_x}$\\
   \textbf{Critic} \textbf{$\mathcal{C}_z$:}\\
   Fix weights $w_{\mathcal{C}_x}$, $w_{\mathcal{E}}$, $w_{\mathcal{D}}$\\
   $ \mathcal{L}_{\mathcal{C}_z}=\frac{1}{b}\sum_{i=1}^{b} \mathcal{C}_z(z_i)-\frac{1}{b}\sum_{i=1}^{b} \mathcal{C}_z(\mathcal{E}(x_i))+\lambda*gp(z_i,\mathcal{E}(x_i))$\\
   $w_{\mathcal{C}_z}=w_{\mathcal{C}_z}-\alpha \sum_{x}\nabla_{w_{\mathcal{C}_z}} \mathcal{L}_{\mathcal{C}_z}$\\
   }
   \textbf{Encoder:}\\
   Fix weights $w_{\mathcal{C}_z}$, $w_{\mathcal{C}_x}$, $w_{\mathcal{D}}$\\
   $ \mathcal{L}_{\mathcal{E}}=\frac{1}{b}\sum_{i=1}^{b} \mathcal{C}_x(x_i)-\frac{1}{b}\sum_{i=1}^{b} \mathcal{C}_x(\mathcal{D}(z_i))+\frac{1}{b}\sum_{i=1}^{b} \mathcal{C}_z(z_i)-\frac{1}{b}\sum_{i=1}^{b} \mathcal{C}_z(\mathcal{E}(x_i))+\sum_{i=1}^{b} {\lVert x_i-\mathcal{D}{(\mathcal{E}(x_i))\rVert}}_2 $\\
   $w_{\mathcal{E}}=w_{\mathcal{E}}-\alpha \sum_{x}\nabla_{w_{\mathcal{E}}} \mathcal{L}_{\mathcal{E}}$\\
   \textbf{Decoder:}\\
   Fix weights $w_{\mathcal{C}_z}$, $w_{\mathcal{C}_x}$, $w_{\mathcal{E}}$\\
   $ \mathcal{L}_{\mathcal{D}}=\frac{1}{b}\sum_{i=1}^{b} \mathcal{C}_x(x_i)-\frac{1}{b}\sum_{i=1}^{b} \mathcal{C}_x(\mathcal{D}(z_i))+\frac{1}{b}\sum_{i=1}^{b} \mathcal{C}_z(z_i)-\frac{1}{b}\sum_{i=1}^{b} \mathcal{C}_z(\mathcal{E}(x_i))+\sum_{i=1}^{b} {\lVert x_i-\mathcal{D}{(\mathcal{E}(x_i))\rVert}}_2 $\\
   $w_{\mathcal{D}}=w_{\mathcal{D}}-\alpha \sum_{x}\nabla_{w_{\mathcal{D}}} \mathcal{L}_{\mathcal{D}}$\\

 }
 \textbf{Testing:}\\
 $X_{test}$=$(x_i^{1...N_w})_{i=1}^n$\\
 \For{$i=1,...,n$}{
    $x_i^{rec}=\mathcal{D}{(\mathcal{E}(x_i))}$\\
    $\mathcal{RE}{(x_i)}=DTW(x_i, x_i^{rec})$\\
    $z_{\mathcal{RE}}{(x_i)}=\frac{\mathcal{RE}{(x_i)}-\mu_{\mathcal{RE}}}{\sigma_{\mathcal{RE}}}$\\
    $z_{\mathcal{CS}}{(x_i)}=\frac{\mathcal{C}_{x}{(x_i)}-\mu_{\mathcal{C}_{x}}}{\sigma_{\mathcal{C}_{x}}}$\\
    $A(x_i)=z_{\mathcal{RE}}{(x_i)}\times z_{\mathcal{CS}}{(x_i)}$

    } 
 \caption{AAE Training and Testing}
\end{algorithm}

\subsection{Parameter Details and Training of AAE}\label{parameters}
The proposed AAE model consists of an autoencoder acting as the generator and two critics of $\mathcal{C}_z$ and $\mathcal{C}_x$ for distinguishing the latent and reconstructed values from the random and real input data, respectively. The input dataset is normalized to be between -1 and 1. To capture the temporal dependency of time-series measurement data, we employ three layers of LSTM with 40 hidden units for the encoder and four layers of LSTM with 40, 80, 40, and 20 hidden units for the decoder, both of which are followed by a dense layer.  We utilize a 1-D convolutional layer for 
$\mathcal{C}_z$ and $\mathcal{C}_x$ followed by a dense layer to recognize the hidden patterns of the reconstructed values.
All the activation functions are ReLU for the autoencoder except the last layer of the decoder, which is Tanh, and LeakyReLU for the critics. Dropout with a rate of 0.2 is also used in every layer of the autoencoder and critics to prevent the model from overfitting. We have performed the \emph{coarse-to-fine} method for optimizing the hyperparameter values by discretizing the values into a grid value and tested the increasing or decreasing their effects on the training results \cite{hyperplane_optimization}. 
The learning rate with the Adam optimizer is initially set to 0.001, multiplied by a decaying factor of 0.99 at each epoch. Another important parameter that has a critical impact on the performance of the training and testing process is window size, which needs to be selected appropriately to capture the attacks' signatures and provide better computational time and accuracy. We use a rolling window strategy with window size 40 and step size 1 to create windows of data points for training and testing the model.
Table \ref{tbl:Hyperparameter_results} shows the validation performance with changes in the number of layers of encoder and decoder, and the changes in the number of neurons within each hidden layer of the autoencoder. Given a large number of possible hyperparameter combinations, only the upper, optimal, and lower bounds were included in the table where increasing (for the upper bound) or decreasing (for the lower bound) the number of layers and neurons had adverse effects on the validation performance.
\subsection{Performance Metrics}\label{Performance_metrics}

To find the reconstruction error ($RE$), dynamic time warping (DTW) \cite{DTW} is applied as a similarity measure, which is more effective in cases of consistent attacks over a period of time. DTW aligns two sequences of time series data by aiming to time the best match between them that represents their similarities. It constructs a distance matrix where its \textit{ij}-th element represents the distance between \textit{i}-th point of the first sequence from the \textit{j}-th point of the second sequence. Consequently, the warping path defines the alignment of the points in the two sequences. Typically, falsified data have higher $RE$ and lower critic scores ($CS$) produced by $C_x$. The final anomaly score associated with these values is calculated as follows
\begin{equation}\label{Anomaly_score}
A(x) = z_{RE}(x) \times z_{CS}(x) 
\end{equation}
where $z_{RE}$ and $z_{CS}$ are the $z$-scores for $RE$ and $CS$, respectively. We use the three-sigma rule to find the threshold of the sliding windows. To evaluate the detection performance, we adopt accuracy ($ACC=(TP+TN)/(TP+TN+FP+FN$) to determine how precise the model can identify both normal and abnormal samples, precision ($Prec=TP/(TP+FP)$) to determine the degree of preciseness in finding anomalies, recall ($Rec=TP/(TP+FN)$) to measure how well the model can detect all ground truths, and F-1 score ($F1=2*(Prec*Rec)/(Prec+Rec$) to compute overall performance of the model using the harmonic mean of precision and recall. Here $TP$ represents true positive, $TN$ is true negative, $FP$ denotes false positive, and $FN$ is false negative.

\begin{table}[t]
    \centering
    \caption{\small  Validation performance of the AAE model for different hyperparameters for IEEE 123-bus.}
      \includegraphics[clip,trim=3.3cm 18.8cm 1.1cm 2.43cm, width=1.15\linewidth]{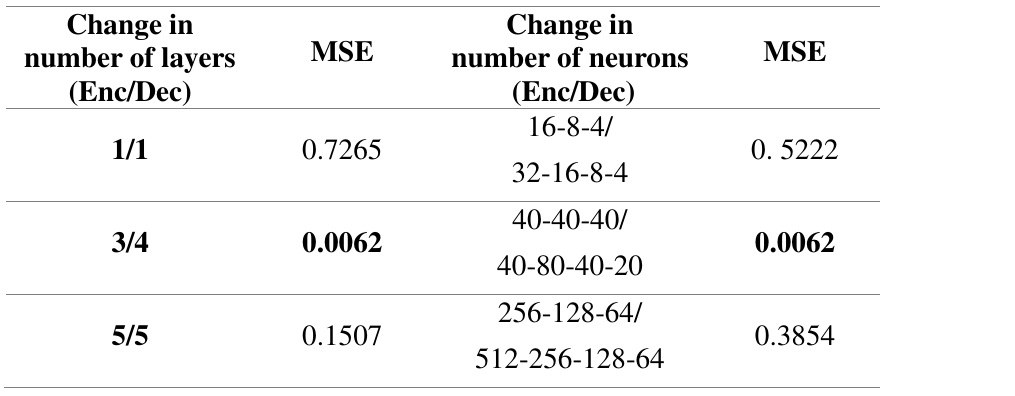}
 \label{tbl:Hyperparameter_results}
 \vspace{-5mm}
\end{table}
\subsection{FDIA Detection Results}The detection performance of the proposed methodology is evaluated in the IEEE 13-bus and IEEE 123-bus distribution grids. All simulation results, including training and testing of the AAE model and other unsupervised models, are conducted with Tensorflow in Python on a computer with Nvidia RTX3070 8GB and a 2.1 GHz Intel Core i7 CPU with 32 GB of RAM. 
\subsubsection{\textbf{IEEE 13-bus System}}
As the sensors are mostly installed on the main buses of the distribution networks, power flow measurements of lines 632-670 in the IEEE 13-bus system are selected for training and applying FDIAs for testing. 80\% of the total 10080 measurements is used for training and 20\% of it for testing. Fig. \ref{loss_13} shows the loss of decoder and critic $\mathcal{C}_x$ during training for 2000 epochs.
As can be seen, at the beginning, the loss of $\mathcal{C}_x$ is very low because it is easy for the critic to recognize the fake values generated by the decoder. However, when the training proceeds, the decoder is trained and able to generate high-quality outputs very close to the original inputs, leading to higher loss values for the critic.

After training the developed AAE model with normal power flow measurements, it is tested under three types of attacks mentioned in Section \ref{FDIA_Development}. Fig. \ref{test_P13} illustrates the AAE model's original input data and reconstructed outputs while the combined FDIA is applied to the test data. As it is shown, when the data is normal, the reconstruction error is very low, meaning that the model is capable of reconstructing the normal data; however, in the presence of attacked data, the model cannot successfully reconstruct the data, causing a large reconstruction error. The detection metrics for these FDIAs are shown in Table \ref{tbl:Performance_Metrics_13}. According to this table, The AAE model has superior performance in detecting all three types of attacks. As the method is reconstructed-based, additive, deductive, or combined functions have similar impacts on the detection performance with a slight difference in the performance metrics. The minor differences in the detection performance show the capability of the proposed AAE model to detect different types of FDIAs successfully.

\subsubsection{\textbf{IEEE 123-bus System}}
This system resembles a large system as a case study, including unbalanced lines and loads. We have integrated solar and wind generation units into this system by installing them randomly on different buses, as shown in Table \ref{tbl:System_Details}. This also incorporates the randomness and uncertainties of DERs into this system, making the proposed methodology more applicable to real-world cases.
\begin{figure}[t!]
    \subfloat[Training loss]{%
        \includegraphics[width=.5\linewidth]{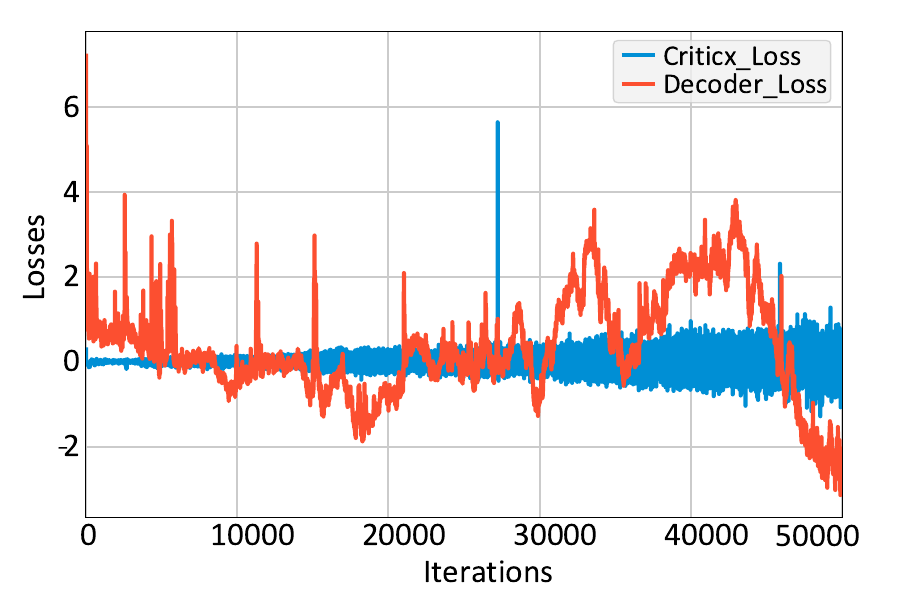}%
        \label{loss_13}%
    }
    \subfloat[Line active power flow]{%
        \includegraphics[width=.5\linewidth]{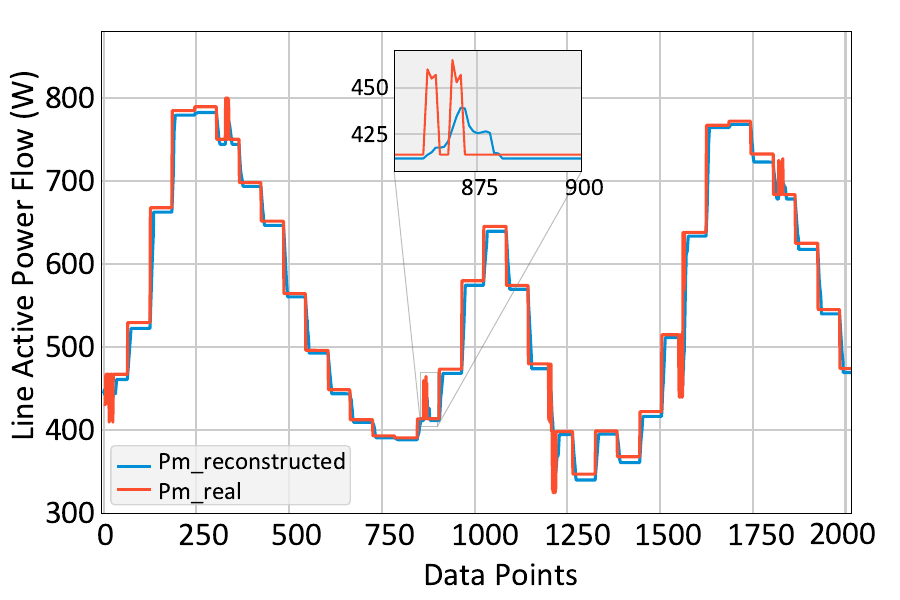}%
        \label{test_P13}%
    }\\
    \subfloat[Training loss]{%
        \includegraphics[width=.5\linewidth]{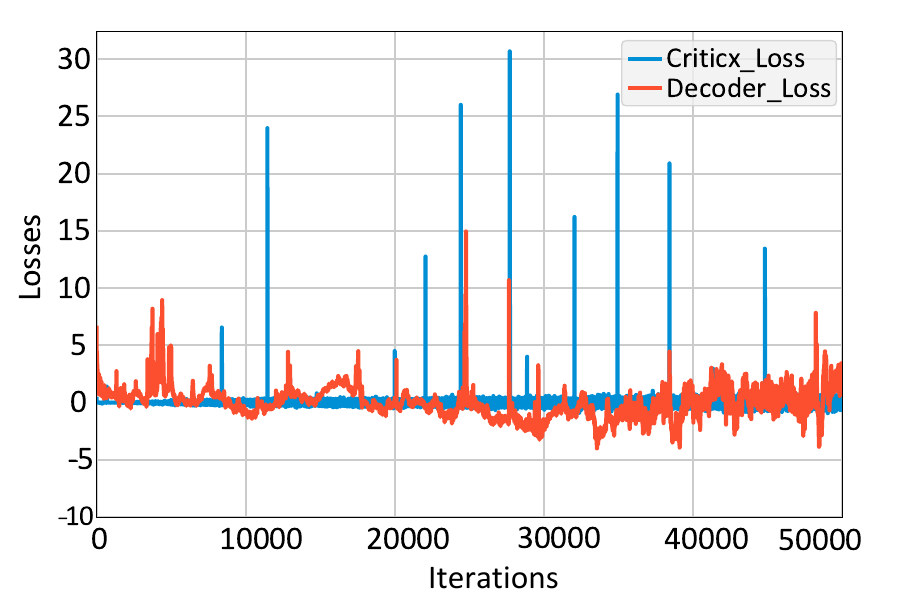}%
        \label{loss_123}%
    }
    \subfloat[Bus voltage magnitude]{%
        \includegraphics[width=.5\linewidth]{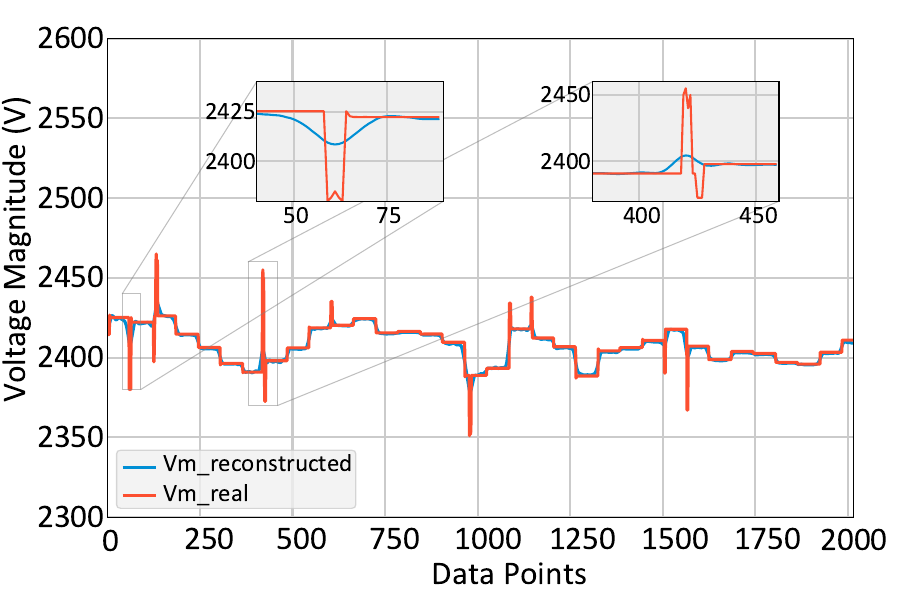}%
        \label{test_V123}%
    }
    \caption{Simulation results of the AAE model for IEEE 13-bus (a and b), and for IEEE 123-bus (c and d).}
    \label{fig:fig}
\end{figure}

To generate simulation data, load and DER profiles spanning one week are employed. In total, 10080 data points are generated, among which 80\% is used for training and 20\% for testing. The systems are monitored in different lines and nodes through $\mu$PMUs to measure the voltage and current magnitudes, voltage and current angles, active power, and reactive power measurements as mentioned in Table \ref{tbl:System_Details}. The training and testing performance details were thoroughly explained in sections \ref{parameters} and \ref{Performance_metrics}. 

The voltage magnitude measurements at bus 35 are used as the train and test datasets. Fig. \ref{loss_123} shows the loss of the decoder(generator) and critic $\mathcal{C}_x$. At the early stages of training, the loss of the critic is very low, indicating that it is easy for the critic to recognize the fake values from the generator. However, while the generator is learning to generate outputs similar to the real data, the loss of the critic increases. To test the proposed model, some parts of the test dataset are falsified using equations \ref{deductive_Eq} and \ref{additive_Eq}.
Fig. \ref{test_V123} shows the output of the AAE model and real test data for the voltage magnitude of bus 35 in the IEEE 123-bus system. As can be seen, in cases of normal data points, the proposed model can reconstruct the data very similarly to the original real values. However, when the data is falsified, the differences between the generated and real values (reconstruction errors) are high. This flags the anomalous data points as attacked data.
The attack detection results of the proposed methodology for the three types of FDIAs are shown in Table \ref{tbl:Performance_Metrics_123}. Similar to the IEEE 13-bus case, the detection accuracy is high, and the recall in all three types of attacks is 95\% and above showing the low rate of FP and high rate of TP. Overall, the high rate of F-1 scores in the three types of attacks, which are more than 90\%, and the high rate of detection accuracy signifies the strong capability of the AAE model in predicting normal and abnormal data points in the whole test dataset. 

\subsection{Comparison with Other Data-Driven Models}
To show the superior performance of the proposed method, its prediction and detection performance is compared with other unsupervised and supervised methods. To make a fair comparison, we use the same training and test strategies of the AAE method, including the same learning rate, window size, step size, dropout, and data normalization strategy as described in sections \ref{parameters} and \ref{Performance_metrics} for reconstruction-based models in sections \ref{AE_LSTM} to \ref{AE_FC}. The details of the data-driven models are described in the following.
\begin{table}[t!]
    \centering
    \caption{\small  Detection metrics of the AAE model under three types of FDIAs for IEEE 13-bus.}
      \includegraphics[clip,trim=4cm 21.3cm 0.9cm 2.53cm, width=1.25\linewidth]{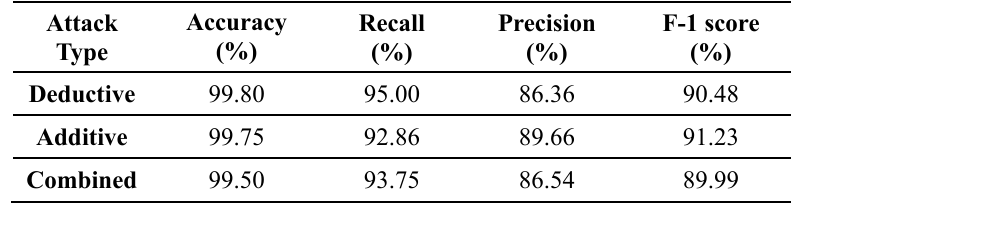}
 
 \label{tbl:Performance_Metrics_13}
\end{table}
\begin{table}[t!]
    \centering
    \caption{\small  Detection metrics of the AAE model under three types of FDIAs for IEEE 123-bus.}
      \includegraphics[clip,trim=4cm 21.3cm 0.9cm 2.53cm, width=1.25\linewidth]{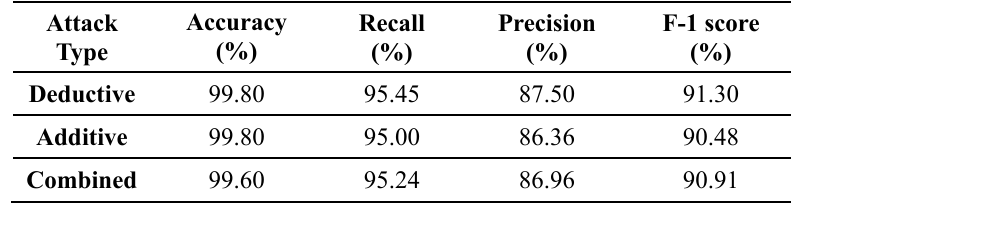}
 \label{tbl:Performance_Metrics_123}
 \vspace{-5mm}
\end{table}
\subsubsection{\textbf{AE-LSTM}}\label{AE_LSTM}
In this model, an autoencoder model with three layers of LSTM with 40 units in the encoder and four layers of LSTM with 40, 80, 40, and 20 hidden units for the decode are chosen. The structure of the autoencoder is the same as that of the AAE model.

\subsubsection{\textbf{AE-CNN}} This model includes two layers of Conv1D with filters of 64 and 32 and kernel sizes 5 and 3, respectively, for the encoder. These layers are followed by one fully connected (FC) layer with a unit number of  20 (latent dimension). For the decoder, the reverse structure of the encoder is applied. Two layers of Conv1D with 32 and 64 filters with kernel sizes 3 and 5, respectively, followed by a fully connected layer with the size of window size.

\subsubsection{\textbf{AE-FC}} \label{AE_FC}
For this model, the encoder consists of three FC layers with 100, 100, and 20 (latent dimension) units. The decoder also has three FC layers with 100, 100, and window size units.

\subsubsection{\textbf{K-Means}} \label{KMeans}
K-Means is a popular unsupervised clustering-based algorithm that partitions a dataset into K separate predefined clusters. For each data point. the distance with its assigned cluster's centroid is calculated and if the calculated distance is greater than a certain threshold, the data point is considered as an anomaly. We have found that for K=32 and K=37, the K-Means algorithm has the best detection performance for 13-bus and 123-bus systems, respectively.

\subsubsection{\textbf{Linear Regression}} \label{LR}
The core idea behind the linear regression (LR) model as a supervised learning model is that it finds the relationships between independent variables and predicted dependent variables using a hyperplane that best fits with the input data. If the residual of the predicted values and input data is beyond the predefined threshold, the values are flagged as anomalies. We have trained the LR model using a sliding window with a window size of 72 and a moving size of 36 to achieve the best performance for both test systems.

\subsubsection{\textbf{OneClassSVM}} \label{SVM}
OneClass support vector machine (SVM) is an unsupervised model that aims to find a soft boundary by transferring the input data to a high-dimensional space using a kernel function. The goal is to maximize the distance from the determined boundary to the origin. We found that the sigmoid kernel function presents the best performance for this model in both test cases.
\vspace{2mm}

\begin{figure}[t!]
    \centering
      \includegraphics[clip,trim=0.51cm 0.2cm -2.6cm 0cm, width=1.25\linewidth]{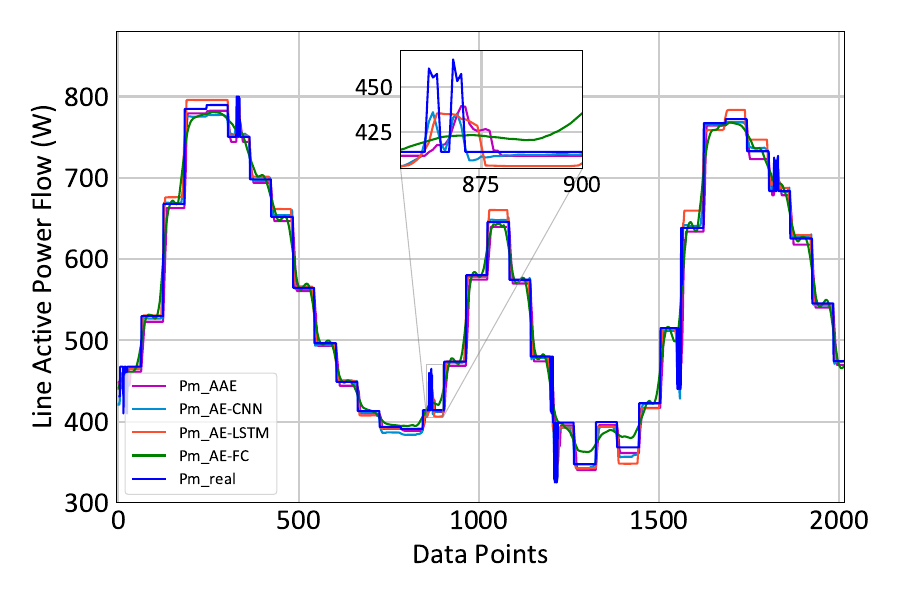}
 \caption{\small  Comparison of the output of the AAE model with outputs of other reconstruction-based models on IEEE 13-bus system test data.}
 \label{fig:Test_P13_comparison}
\end{figure}

Fig. \ref{fig:Test_P13_comparison} shows the outputs of the proposed AAE  and other unsupervised reconstruction-based methods under combined attack for the IEEE 13-bus system. As shown in this figure and by comparing the performance metrics of the methods in Table \ref{tbl:Performance_Comparison_13}, it is evident that the recall rates of AE-LSTM and AE-CNN are close to that of AAE whereas there is a large difference between the precision rates for AE-LSTM and AE-CNN (73.33\% and 69.84\%), and precision of the AAE model (86.54\%) due to large number of FP in the detection of the former methods. This leads to lower rates of F-1 scores for these methods compared to the proposed AAE model. The performance metrics of the AE-FC model are lower than the other three models because this model, as shown in Fig. \ref{fig:Test_P13_comparison} cannot accurately reconstruct the data points causing higher numbers of FP and FN. The proposed AAE model also outperforms K-Means, OneclassSVM, and LR models in terms of accuracy, recall, and F-1 score. Among these three models, K-Means shows the highest performance by having an F-1 score of 86.60\%. Although OneClassSVM does not have any FP, it has the poorest performance due to a low TP rate.

\begin{figure}[t!]
    \centering
    \includegraphics[clip,trim=0.51cm 0.2cm -2.6cm 0cm, width=1.25\linewidth]{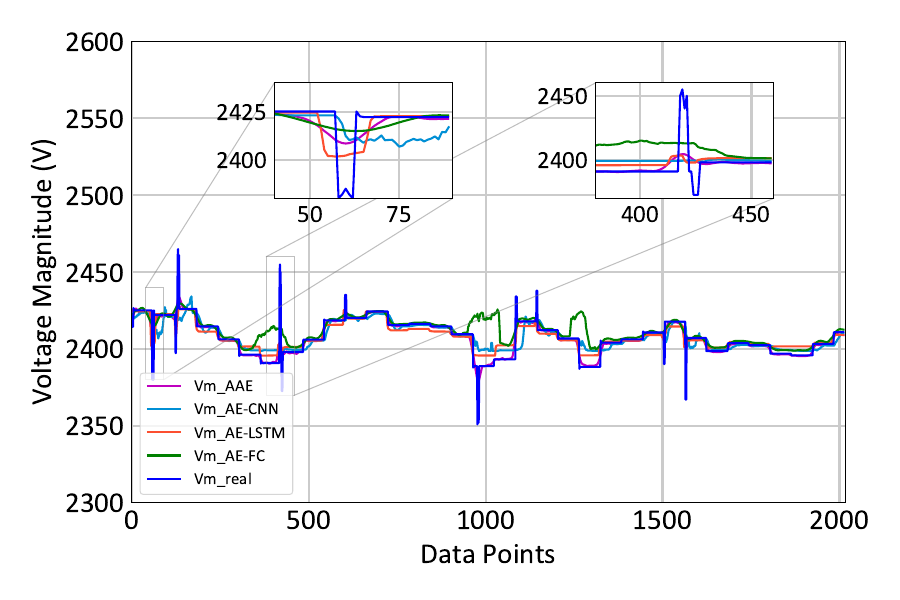}
 \caption{\small  Comparison of the output of the AAE model with outputs of other reconstruction-based models on IEEE 123-bus system test data.}
 \label{fig:Test_V123_comparison}
\end{figure}

The prediction results of the AAE model and other unsupervised reconstruction-based methods under the combined attack for the IEEE 123-bus system are shown in \ref{fig:Test_V123_comparison}. Table \ref{tbl:Performance_Comparison_123} also summarizes the performance metrics of the AAE and other data-driven methods for the IEEE 123-bus network. Following the figures and metrics, the AAE performs significantly better than the other reconstruction-based methods. The results of AE-LSTM and AE-CNN are very similar to each other. Although these models have a high rate of recall, the high number of FP decreases their precision (67.86\% and 66.10\%), causing lower rates of F-1 scores (77.55\% and 77.23\%). On the other hand, when AE-FC is employed, due to the lower capability of the network to reconstruct the normal data and identify the abnormal points, the recall rate is drastically decreased by 35.72\%. 

\begin{table}[h]
    \centering
    \caption{\small  Performance comparison under combined attack for IEEE 13-bus.}
      \includegraphics[clip,trim=3.5cm 18.5cm 0.0cm 2.53cm, width=1.25\linewidth]{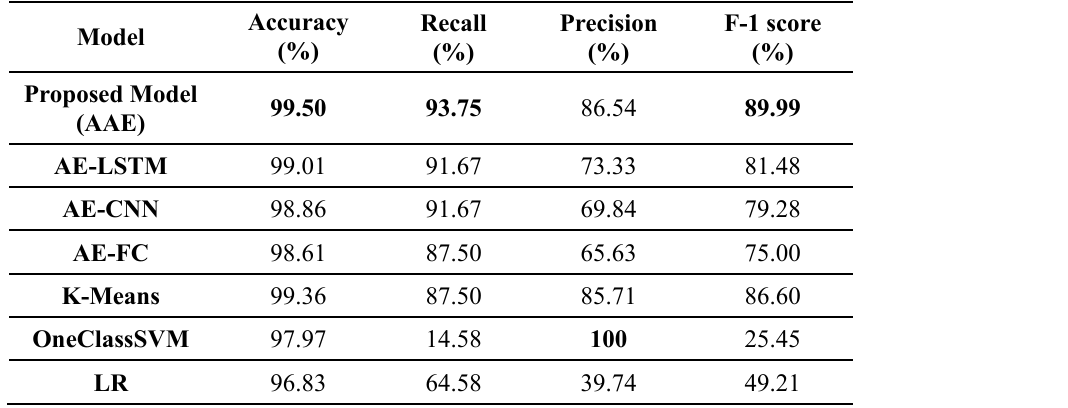}
 
 \label{tbl:Performance_Comparison_13}
\end{table}
This is because of a lower rate of TP and a higher rate of FN compared to the AAE model. Compared with K-Means, OneClassSVM, and LR, the proposed AAE model has the best performance with accuracy and F-1 score of 99.60\% and 90.91\%, respectively. LR had the second-best performance with the same accuracy as the AAE model and an F-1 score of 89.47\%. It did not detect any FP, and so did OneClassSVM. However, the poorest performance of OneClassSVM originates from the low TP rate.

The simulation results of the AAE model and the comparison with other data-driven methods signify three important aspects of this model. Firstly, the GAN model embedded in the AAE model can better reconstruct the normal data and provide higher reconstruction errors for anomalous points. Secondly, using the combination of reconstruction error and the critic score leads to higher anomaly scores and performance metrics compared to other unsupervised methods, which only use the reconstruction errors. 
Lastly, as shown in Tables \ref{tbl:Performance_Comparison_13} and \ref{tbl:Performance_Comparison_123}, the performance of other models especially regression and clustering models significantly changes when they are tested on different test systems whereas since the proposed AAE model can successfully model the nonlinear patterns of the input data, it has the highest performance in both test systems. While the proposed AAE model requires a longer training time compared to other supervised and unsupervised models due to the embedded GAN and AE components, once it is trained on the historical training data, it can be deployed to the systems with significantly low detection time. According to our simulations, once the initial training is performed, it takes approximately 96  and 98 milliseconds to determine the new incoming data as anomalous or normal for 13-bus and 123-bus systems, respectively. 
\begin{table}[t!]
    \centering
    \caption{\small  Performance comparison under combined attack for IEEE 123-bus.}
      \includegraphics[clip,trim=3.5cm 18.5cm 0.0cm 2.53cm, width=1.25\linewidth]{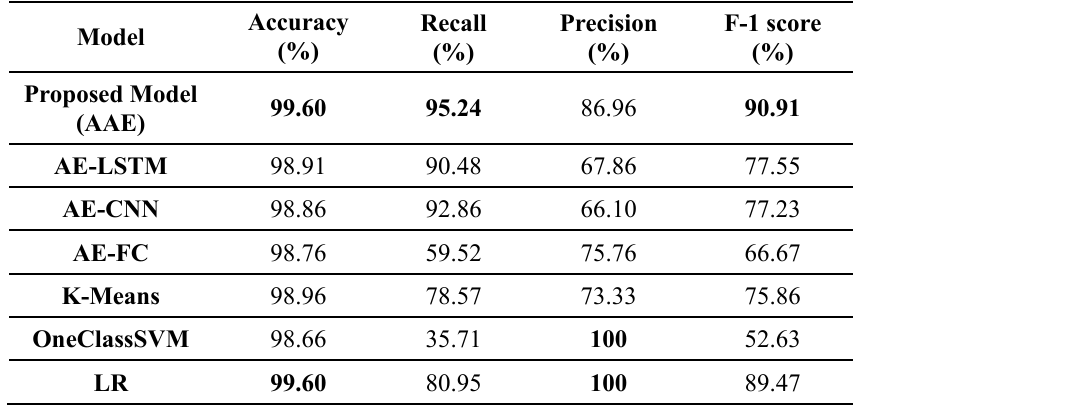}
\label{tbl:Performance_Comparison_123}
\end{table}
\section{Conclusion}\label{conclusion}
This paper proposed an unsupervised method for FDIA detection in unbalanced power distribution grids. The proposed AAE learning-based approach employs LSTM in the architecture of the autoencoder for capturing the temporal correlation of the grid measurements to reconstruct the input measurement samples and uses two critics to distinguish the generated samples from the real data points. In order to identify the anomalous points, the reconstruction errors and outputs of the critic were fused to create a combined anomaly score. The performance of the proposed approach was evaluated on IEEE 13-bus and IEEE 123-bus systems with historical load data and meteorological data for DGs. The performance metrics of the proposed method were compared with those of other unsupervised methods. The experimental results showed the superior performance of the proposed method under three types of FDIAs without the need for labeled data due to the use of the GAN model and the combined anomaly score. In addition, the AAE model outperformed other unsupervised methods by having higher detection metrics in both standard test systems. In our future work, the dynamics of power distribution systems and inverter-based resources will be analyzed to develop physics-informed machine learning models by integrating the underlying physical principles into ML algorithms for cyber-physical event analysis.




\bibliographystyle{ieeetr}
\bibliography{EPSR_AAE.bib}

\begin{thebibliography}{10}

\bibitem{protocols}
M.~Hasan, A.~Habib, Z.~Shukur, F.~Ibrahim, S.~Islam, and M.~Razzaque, ``Review on cyber-physical and cyber-security system in smart grid: Standards, protocols, constraints, and recommendations,'' {\em Journal of Network and Computer Applications}, vol.~209, p.~103540, Jan 2023. https://doi.org/10.1016/j.jnca.2022.103540.

\bibitem{cyber-physical}
S.~Majumder, A.~Vosughi, H.~Mustafa, T.~Warner, and A.~Srivastava, ``On the cyber-physical needs of der-based voltage control/optimization algorithms in active distribution network,'' {\em IEEE Access}, May 2023. https://doi.org/10.1109/ACCESS.2023.3278281.

\bibitem{Dileep}
G.~Dileep, ``A survey on smart grid technologies and applications,'' {\em Renewable energy}, vol.~146, pp.~2589--2625, Feb 2020. https://doi.org/10.1016/j.renene.2019.08.092.

\bibitem{mirzapour_GET}
O.~Mirzapour, X.~Rui, and M.~Sahraei-Ardakani, ``Grid-enhancing technologies: Progress, challenges, and future research directions,'' {\em Electric Power Systems Research}, vol.~230, p.~110304, 2024.

\bibitem{sparsity_con}
X.~Ye, I.~Esnaola, S.~Perlaza, and R.~Harrison, ``Stealth data injection attacks with sparsity constraints,'' {\em IEEE Transactions on Smart Grid}, Jan 2023. https://doi.org/10.1109/TSG.2023.3238913.

\bibitem{observer_bonab}
P.~A. Bonab, J.~Holland, and A.~Sargolzaei, ``An observer-based control for a networked control of permanent magnet linear motors under a false-data-injection attack,'' in {\em 2023 IEEE Conference on Dependable and Secure Computing (DSC)}, pp.~1--8, 2023.

\bibitem{vincent_cyberphysical}
E.~Vincent, M.~Korki, M.~Seyedmahmoudian, A.~Stojcevski, and S.~Mekhilef, ``Detection of false data injection attacks in cyber-physical systems using graph convolutional network,'' {\em Electric Power Systems Research}, vol.~217, p.~109118, Apr 2023.

\bibitem{reachability_FDIA}
R.~Liu, H.~Mustafa, Z.~Nie, and A.~Srivastava, ``Reachability-based false data injection attacks and defence mechanisms for cyberpower system,'' {\em Energies}, vol.~15, no.~5, p.~1754, Feb 2022. https://doi.org/10.3390/en15051754.

\bibitem{masking}
X.~Liu, L.~Zhiyi, L.~Xingdong, and L.~Zuyi, ``Masking transmission line outages via false data injection attacks,'' {\em IEEE Transactions on Information Forensics and Security}, vol.~11, no.~7, pp.~1592--1602, Mar 2016. https://doi.org/10.1109/TIFS.2016.2542061.

\bibitem{survey_physics}
J.~Giraldo, D.~Urbina, A.~Cardenas, J.~Valente, M.~Faisal, J.~Ruths, N.~Tippenhauer, H.~Sandberg, and R.~Candell, ``A survey of physics-based attack detection in cyber-physical systems,'' {\em ACM Computing Surveys (CSUR)}, vol.~51, no.~4, pp.~1--36, Jul 2018. https://doi.org/10.1145/3203245.

\bibitem{model_based_AGC}
S.~Sridhar and M.~Govindarasu, ``Model-based attack detection and mitigation for automatic generation control,'' {\em IEEE Transactions on Smart Grid}, vol.~5, no.~2, pp.~580--591, Mar 2014. https://doi.org/10.1109/TSG.2014.2298195.

\bibitem{survey_fdia}
A.~S. Musleh, C.~Guo, and Y.~D. Zhao, ``A survey on the detection algorithms for false data injection attacks in smart grids,'' {\em IEEE Transactions on Smart Grid}, vol.~11, no.~3, pp.~2218--2234, Oct 2019. https://doi.org/10.1109/TSG.2019.2949998.

\bibitem{yashar_datadriven}
A.~Mehrzad, M.~Darmiani, Y.~Mousavi, M.~Shafie-Khah, and M.~Aghamohammadi, ``A review on data-driven security assessment of power systems: Trends and applications of artificial intelligence,'' {\em IEEE Access}, vol.~11, pp.~78671--78685, 2023.

\bibitem{PIML}
M.~J. Zideh, P.~Chatterjee, and A.~K. Srivastava, ``Physics-informed machine learning for data anomaly detection, classification, localization, and mitigation: A review, challenges, and path forward,'' {\em IEEE Access}, vol.~12, pp.~4597--4617, 2024.

\bibitem{PIConvAE}
M.~J. Zideh and S.~K. Solanki, ``Physics-informed convolutional autoencoder for cyber anomaly detection in power distribution grids,'' {\em arXiv preprint arXiv:2312.04758}, Dec 2023.

\bibitem{graph_FDIA}
X.~Li, Y.~Wang, and Z.~Lu, ``Graph-based detection for false data injection attacks in power grid,'' {\em Energy}, vol.~263, p.~125865, Jan 2023. https://doi.org/10.1016/j.energy.2022.125865.

\bibitem{DL_state_prediction}
H.~Wang, J.~Ruan, Z.~Ma, B.~Zhou, X.~Fu, and G.~Cao, ``Deep learning aided interval state prediction for improving cyber security in energy internet,'' {\em Energy}, vol.~174, pp.~1292--1304, May 2019. https://doi.org/10.1016/j.energy.2019.03.009.

\bibitem{Data_Integrity}
H.~Goyel and S.~K.~Shanti, ``Data integrity attack detection using ensemble based learning for cyber physical power systems,'' {\em IEEE Transactions on Smart Grid}, Aug 2022. https://doi.org/10.1109/TSG.2022.3199305.

\bibitem{multivariat_eensemble}
Y.~Li, W.~Xue, T.~Wu, H.~Wang, B.~Zhou, S.~Aziz, and Y.~He, ``Intrusion detection of cyber physical energy system based on multivariate ensemble classification,'' {\em Energy}, vol.~218, p.~119505, Mar 2021. https://doi.org/10.1016/j.energy.2020.119505.

\bibitem{realtime}
Y.~He, J.~M. Gihan, and W.~Jin, ``Real-time detection of false data injection attacks in smart grid: A deep learning-based intelligent mechanism,'' {\em IEEE Transactions on Smart Grid}, vol.~8, no.~5, pp.~2505--2516, May 2017. https://doi.org/10.1109/TSG.2017.2703842.

\bibitem{inertia_recovery}
J.~Ruan, G.~Liang, J.~Zhao, J.~Qiu, and Z.~Y. Dong, ``An inertia-based data recovery scheme for false data injection attack,'' {\em IEEE Transactions on Industrial Informatics}, vol.~18, no.~11, pp.~7814--7823, 2022.

\bibitem{generalized_recovery}
Y.~Zhu, J.~Ruan, G.~Fan, S.~Wang, G.~Liang, and J.~Zhao, ``A generalized data recovery model against false data injection attack in smart grid,'' in {\em 2022 IEEE 6th Conference on Energy Internet and Energy System Integration (EI2)}, pp.~1477--1482, 2022.

\bibitem{gan_recovery}
Y.~Li, Y.~Wang, and S.~Hu, ``Online generative adversary network based measurement recovery in false data injection attacks: A cyber-physical approach,'' {\em IEEE Transactions on Industrial Informatics}, vol.~16, no.~3, pp.~2031--2043, 2020.

\bibitem{DL_cyber_review}
J.~Ruan, G.~Liang, J.~Zhao, H.~Zhao, J.~Qiu, F.~Wen, and Z.~Y. Dong, ``Deep learning for cybersecurity in smart grids: Review and perspectives,'' {\em Energy Conversion and Economics}, vol.~4, no.~4, pp.~233--251, 2023.

\bibitem{DG_impact}
S.~Razavi, E.~Rahimi, M.~Javadi, A.~Nezhad, M.~Lotfi, M.~Shafie-khah, and J.~Catalão, ``Impact of distributed generation on protection and voltage regulation of distribution systems: A review,'' {\em Renewable and Sustainable Energy Reviews}, vol.~105, pp.~157--167, May 2019. https://doi.org/10.1016/j.rser.2019.01.050.

\bibitem{pf_tiwari}
D.~Tiwari, M.~J. Zideh, V.~Talreja, V.~Verma, S.~K. Solanki, and J.~Solanki, ``Power flow analysis using deep neural networks in three-phase unbalanced smart distribution grids,'' {\em IEEE Access}, vol.~12, pp.~29959--29970, 2024.

\bibitem{transfer_hai}
A.~Hai, T.~Dokic, M.~Pavlovski, T.~Mohamed, D.~Saranovic, M.~Alqudah, M.~Kezunovic, and Z.~Obradovic, ``Transfer learning for event detection from pmu measurements with scarce labels,'' {\em IEEE Access}, vol.~9, pp.~127420--127432, Sep 2021. https://doi.org/10.1109/ACCESS.2021.3111727.

\bibitem{AD_building}
C.~Fan, F.~Xiao, Y.~Zhao, and J.~Wang, ``Analytical investigation of autoencoder-based methods for unsupervised anomaly detection in building energy data,'' {\em Applied Energy}, vol.~211, pp.~1123--1135, Feb 2018. https://doi.org/10.1016/j.apenergy.2017.12.005.

\bibitem{aligholian2021}
A.~Aligholian, A.~Shahsavari, E.~Stewart, E.~Cortez, and H.~Mohsenian-Rad, ``Unsupervised event detection, clustering, and use case exposition in micro-pmu measurements,'' {\em IEEE Transactions on Smart Grid}, vol.~12, no.~4, pp.~3624--3636, Mar 2021. https://doi.org/10.1109/TSG.2021.3063088.

\bibitem{fast_ramped}
N.~Müller, C.~Heinrich, K.~Heussen, and H.~Bindner, ``Unsupervised detection and open-set classification of fast-ramped flexibility activation events,'' {\em Applied Energy}, vol.~312, p.~118647, Apr 2022. https://doi.org/10.1016/j.apenergy.2022.118647.

\bibitem{aligholian2020}
A.~Aligholian, A.~Shahsavari, E.~Cortez, E.~Stewart, and H.~Mohsenian-Rad, ``Event detection in micro-pmu data: A generative adversarial network scoring method,'' in {\em 2020 IEEE Power \& Energy Society General Meeting (PESGM)}, pp.~1--5, Aug 2020. https://doi.org/10.1109/PESGM41954.2020.9281560.

\bibitem{solar_farm}
M.~Dey, S.~Rana, C.~Simmons, and S.~Dudley, ``Solar farm voltage anomaly detection using high-resolution $micro$pmu data-driven unsupervised machine learning,'' {\em Applied Energy}, vol.~303, p.~117656, Dec 2021. https://doi.org/10.1016/j.apenergy.2021.117656.

\bibitem{david2020}
D.~Amoateng, R.~Yan, and T.~Saha, ``A deep unsupervised learning approach to pmu event detection in an active distribution network,'' in {\em 2020 IEEE Power \& Energy Society General Meeting (PESGM)}, pp.~1--5, Aug 2020. https://doi.org/10.1109/PESGM41954.2020.9281767.

\bibitem{semi-Supervised}
Y.~Zhang, W.~Jianhui, and C.~Bo, ``Detecting false data injection attacks in smart grids: A semi-supervised deep learning approach,'' {\em IEEE Transactions on Smart Grid}, vol.~12, no.~1, pp.~623--634, Jul 2020. https://doi.org/10.1109/TSG.2020.3010510.

\bibitem{hierarchical}
Q.~Li, Z.~Jinan, Z.~Junbo, Y.~Jin, S.~Wenzhan, and L.~Fangyu, ``Adaptive hierarchical cyber attack detection and localization in active distribution systems,'' {\em IEEE transactions on smart grid}, vol.~13, no.~3, pp.~2369--2380, Feb 2022. https://doi.org/10.1109/TSG.2022.3148233.

\bibitem{volt_reg}
N.~Bhusal, G.~Mukesh, and B.~Mohammed, ``Detection of cyber attacks on voltage regulation in distribution systems using machine learning,'' {\em IEEE Access}, vol.~9, pp.~40402--40416, Mar 2021. https://doi.org/10.1109/ACCESS.2021.3064689.

\bibitem{coordinated}
N.~Bhusal, M.~Gautam, R.~Shukla, M.~Benidris, and S.~Sengupta, ``Coordinated data falsification attack detection in the domain of distributed generation using deep learning,'' {\em International Journal of Electrical Power \& Energy Systems}, vol.~134, p.~107345, Jan 2022. https://doi.org/10.1016/j.ijepes.2021.107345.

\bibitem{naderi2022}
E.~Naderi, A.~Aydeger, and A.~Asrari, ``Detection of false data injection cyberattacks targeting smart transmission/distribution networks,'' in {\em 2022 IEEE Conference on Technologies for Sustainability (SusTech)}, pp.~224--229, Apr 2022. https://doi.org/10.1109/SusTech53338.2022.9794237.

\bibitem{raghuvamsi}
Y.~Raghuvamsi and K.~Teeparthi, ``Detection and reconstruction of measurements against false data injection and dos attacks in distribution system state estimation: A deep learning approach,'' {\em Measurement}, vol.~210, p.~112565, Mar 2023. https://doi.org/10.1016/j.measurement.2023.112565.

\bibitem{radhoush2023}
S.~Radhoush, T.~Vannoy, K.~Liyanage, B.~Whitaker, and H.~Nehrir, ``Distribution system state estimation and false data injection attack detection with a multi-output deep neural network,'' {\em Energies}, vol.~16, no.~5, p.~2288, Feb 2023. https://doi.org/10.3390/en16052288.

\bibitem{super_resolution_perception}
J.~Ruan, G.~Fan, Y.~Zhu, G.~Liang, J.~Zhao, F.~Wen, and Z.~Y. Dong, ``Super-resolution perception assisted spatiotemporal graph deep learning against false data injection attacks in smart grid,'' {\em IEEE Transactions on Smart Grid}, vol.~14, no.~5, pp.~4035--4046, 2023.

\bibitem{Conv_AE_ehsani}
N.~Ehsani, F.~Aminifar, and H.~Mohsenian‐Rad, ``Convolutional autoencoder anomaly detection and classification based on distribution pmu measurements,'' {\em IET Generation, Transmission \& Distribution}, vol.~16, no.~14, pp.~2816--2828, 2022.

\bibitem{camouflage}
S.~Bhattacharjee, A.~Thakur, and S.~Das, ``Towards fast and semi-supervised identification of smart meters launching data falsification attacks,'' in {\em Proceedings of the 2018 on Asia Conference on Computer and Communications Security}, pp.~173--185, May 2018. https://doi.org/10.1145/3196494.3196551.

\bibitem{solar_irad}
R.~Nematirad and A.~Pahwa, ``Solar radiation forecasting using artificial neural networks considering feature selection,'' in {\em 2022 IEEE Kansas Power and Energy Conference (KPEC)}, pp.~1--4, Apr 2022. https://doi.org/10.1109/KPEC54747.2022.9814765.

\bibitem{bhanja}
S.~Bhanja and A.~Das, ``Impact of data normalization on deep neural network for time series forecasting,'' {\em arXiv preprint arXiv:1812.05519}, Dec 2018. https://doi.org/10.48550/arXiv.1812.05519.

\bibitem{sliding_window}
H.~Saeed, H.~Wang, M.~Peng, A.~Hussain, and A.~Nawaz, ``Online fault monitoring based on deep neural network \& sliding window technique,'' {\em Progress in Nuclear Energy}, vol.~121, p.~103236, Mar 2020. https://doi.org/10.1016/j.pnucene.2019.103236.

\bibitem{stock_idex}
S.~Nayak, B.~Misra, and H.~Behera, ``Impact of data normalization on stock index forecasting,'' {\em International Journal of Computer Information Systems and Industrial Management Applications}, vol.~6, no.~2014, pp.~257--269, Jan 2014.

\bibitem{nemati_acoustic}
R.~Nematirad, M.~Behrang, and A.~Pahwa, ``Acoustic-based online monitoring of cooling fan malfunction in air-forced transformers using learning techniques,'' {\em IEEE Access}, vol.~12, pp.~26384--26400, 2024.

\bibitem{jayala}
T.~Jayalakshmi and A.~Santhakumaran, ``Statistical normalization and back propagation for classification,'' {\em International Journal of Computer Theory and Engineering}, vol.~3, no.~1, pp.~1793--8201, Feb 2011. https://doi.org/10.7763/IJCTE.2011.V3.288.

\bibitem{TadGAN}
A.~Geiger, D.~Liu, S.~Alnegheimish, A.~Cuesta-Infante, and K.~Veeramachaneni, ``Tadgan: Time series anomaly detection using generative adversarial networks,'' in {\em 2020 IEEE International Conference on Big Data (Big Data)}, pp.~33--43, Dec 2020. https://doi.org/10.1109/BigData50022.2020.9378139.

\bibitem{autoencoder_AD}
M.~Sakurada and T.~Yairi, ``Anomaly detection using autoencoders with nonlinear dimensionality reduction,'' in {\em 2Proceedings of the MLSDA 2014 2nd workshop on machine learning for sensory data analysis}, pp.~4--11, Dec 2014. http://dx.doi.org/10.1145/2689746.2689747.

\bibitem{AD_Yin}
C.~Yin, S.~Zhang, J.~Wang, and N.~N. Xiong, ``Anomaly detection based on convolutional recurrent autoencoder for iot time series,'' {\em IEEE Transactions on Systems, Man, and Cybernetics: Systems}, vol.~52, no.~1, pp.~112--122, 2022.

\bibitem{iterative}
S.~S. Mohtavipour and M.~Jabbari~Zideh, ``An iterative method for detection of the collusive strategy in prisoner’s dilemma game of electricity market,'' {\em EEJ Transactions on Electrical and Electronic Engineering}, vol.~14, no.~2, pp.~252--260, Feb 2019. https://doi.org/10.1002/tee.22804.

\bibitem{GAN}
I.~Goodfellow, J.~Pouget-Abadie, M.~Mirza, B.~Xu, D.~Warde-Farley, S.~Ozair, A.~Courville, and Y.~Bengio, ``Generative adversarial networks,'' {\em arXiv preprint arXiv:1406.2661}, 2014. https://doi.org/10.48550/arXiv.1406.2661.

\bibitem{WGAN}
M.~Arjovsky, S.~Chintala, and L.~Bottou, ``Wasserstein generative adversarial networks,'' in {\em International conference on machine learning}, pp.~214--223, PMLR, Jul 2017.

\bibitem{gradient}
I.~Gulrajani, F.~Ahmed, M.~Arjovsky, V.~Dumoulin, and A.~Courville, ``Improved training of wasserstein gans,'' in {\em Proc. of the 31st Int. Conf. on Neural Information Processing Systems}, p.~5769–5779, 2017.

\bibitem{test_systems}
``Ieee pes distribution systems analysis subcommittee radial test feeders. [online]. available: https://cmte.ieee.org/pes-testfeeders/resources/. accessed: Sep 2023,''

\bibitem{NREL}
 “Weather Data,” National Renewable Energy Laboratory, [Online]. Available: https://sam.nrel.gov/weather-data.html. [Accessed 5 Dec 2022].

\bibitem{hyperplane_optimization}
J.~Bergstra, R.~Bardenet, Y.~Bengio, and B.~Kégl, ``Algorithms for hyper-parameter optimization,'' in {\em Advances in neural information processing systems}, p.~2546–2554, 2011.

\bibitem{DTW}
D.~Berndt and J.~Clifford, ``Using dynamic time warping to find patterns in time series,'' in {\em KDD workshop}, pp.~359--370, Jul 1994.

\end{thebibliography}


\end{document}